\documentclass[a4paper,10pt]{article}
\usepackage[utf8]{inputenc}
\usepackage{amsmath}
\usepackage{mathtools}
\usepackage{amsfonts}
\usepackage{graphicx}
\usepackage{epstopdf}
\usepackage{caption}
\usepackage{hyperref}
\usepackage{subcaption}
\usepackage{lscape}
\usepackage{xcolor}
\usepackage{multirow}
\usepackage{cellspace}
\usepackage{bbm}
\usepackage{rotating}

\usepackage[toc]{appendix}

\DeclareMathOperator*{\argmax}{argmax}

\usepackage[a4paper, total={7in, 10in}]{geometry}

\usepackage{authblk}
\usepackage[authoryear]{natbib}

\author[1]{Giuseppe Buccheri}
\author[2]{Giacomo Bormetti}
\author[3,4]{Fulvio Corsi}
\author[2,5]{Fabrizio Lillo}
\affil[1]{\small{Scuola Normale Superiore, Italy}}
\affil[2]{\small{University of Bologna, Italy}}
\affil[3]{\small{University of Pisa, Italy}} 
\affil[4]{\small{City University of London, UK}}
\affil[5]{\small{CADS, Human Technopole, Milan, Italy}}

\renewcommand{\baselinestretch}{1.5}

\begin{document}


\title{A Score-Driven Conditional Correlation Model for Noisy and Asynchronous Data: an Application to High-Frequency Covariance Dynamics\footnote{Corresponding author: \href{mailto:fabrizio.lillo@unibo.it}{fabrizio.lillo@unibo.it}. We are particularly grateful for suggestions we have received from Maria Elvira Mancino, Davide Delle Monache, Ivan Petrella, Fabrizio Venditti, Giampiero Gallo, Davide Pirino and participants to the IAAE 2017 conference in Sapporo, the 10$^{\text{th}}$ SoFiE conference in New York and the VIECO 2017 conference in Wien.}}

\date{}
\maketitle
\begin{center}
First version: December, 2017\\
This version: March, 2019
\end{center}

\begin{abstract}

The analysis of the intraday dynamics of correlations among high-frequency returns is challenging due to the presence of asynchronous trading and market microstructure noise. Both effects may lead to significant data reduction and may severely underestimate correlations if traditional methods for low-frequency data are employed. We propose to model intraday log-prices through a multivariate local-level model with score-driven covariance matrices and to treat asynchronicity as a missing value problem. The main advantages of this approach are: (i) all available data are used when filtering correlations, (ii) market microstructure noise is taken into account, (iii) estimation is performed through standard maximum likelihood methods. Our empirical analysis, performed on 1-second NYSE data, shows that opening hours are dominated by idiosyncratic risk and that a market factor progressively emerges in the second part of the day. The method can be used as a nowcasting tool for high-frequency data, allowing to study the real-time response of covariances to macro-news announcements and to build intraday portfolios with very short optimization horizons. 

 
\vspace{1cm}

\end{abstract}

\newpage

\section{Introduction} 
\label{sec:intro}

A large class of conditional covariance models have been proposed in the econometric literature and their use is widespread in risk and portfolio management at daily or lower frequencies. Popular multivariate dynamic time-series models include the class of multivariate extensions of the univariate GARCH model of \cite{Engle} and \cite{Bollerslev} and the Dynamic Conditional Correlation (DCC) model of \cite{EngleDCC}. A drawback of these models is that they are misspecified if data are recorded with observational noise and require synchronization in case data are irregularly spaced. As a consequence, they cannot be straightforwardly applied to intraday data, since high-frequency prices are contaminated by microstructure noise and assets are traded asynchronously. Both effects may lead to ignore a large portion of data and can significantly underestimate correlations. The problem of estimating and forecasting intraday volatilities and correlations is, however, of crucial importance in high-frequency finance. For instance, an high-frequency trader is interested in rebalancing the portfolio on an intraday basis and thus needs accurate short-term covariance forecasts. Similarly, the study of the intraday dependencies of financial assets is useful to examine the reaction of the market to external information and has a theoretical relevance in market microstructure research. 

We contribute to the literature on intraday covariance estimation by proposing a modelling strategy that can handle both asynchronous trading and microstructure effects. High-frequency log-prices are modeled through a conditionally norma local-level model where efficient log-prices are affected by measurement errors and the covariance matrices of both the efficient returns and the noise are time-varying. In this state-space representation, asynchronous trading can be treated as a standard missing value problem. The dynamics of time-varying parameters are driven by the score of the conditional density (\citealt{GAS1}, \citealt{Harvey_2013}). The latter can easily be computed using the standard Kalman filter, as described by \cite{GAS4} and \cite{DelleMonache2}.  

The main advantage of this state-space representation is that it allows to model the covariances of latent efficient returns using all available observed prices. In standard conditional correlation models, the covariances of observed returns are instead modelled. While in a low-frequency setting the two approaches are equivalent, at high-frequency, models for observed returns are subject to large data reduction and are affected by microstructure noise. For instance, assume the $i$-th asset is traded at time $t$, but is not traded at time $t-1$, a circumstance that is very common in practice. The local-level model can exploit the observation of the price of the $i$-th asset at time $t$ to reconstruct the efficient price and to update correlations. In contrast, in standard conditional correlation models, the return of the $i$-th asset at time $t$ is treated as a missing value and the information related to the price of the traded asset is neglected. An alternative method would be synchronizing the data, for instance by previous-tick interpolation. This leads to a large number of zero returns, which are known to jeopardize the inference. This effect is analogous to the downward bias of high-frequency sample correlations, the well-known ``Epps effect'' (\citealt{Epps}), which arises when using previous-tick or other interpolation schemes (see \citealt{HayashiYoshida} and references therein). Microstructure effects constitute an additional source of bias for standard conditional correlation models. Another immediate consequence of modelling the covariances of latent efficient returns is that correlation estimates are robust to measurement errors. Compared to standard dynamic covariance models, the proposed approach is thus specifically designed to deal with high-frequency data and can easily be employed to construct intraday portfolios. To this end, we discuss two alternative parameterizations of the covariance matrix leading to positive-definite estimates.

Since \cite{AndersenBollerslev} and \cite{Tsay}, it is known that intraday volatilities have the typical U-shape, being larger at the opening and closing hours of the trading day. In contrast, due to the aforementioned difficulties, the intraday behavior of correlations has received less attention in the financial econometric literature. Notably exceptions are given by the work of \cite{BibingerSC}, who proposed a nonparametric spot covariance estimator for intraday data and those of \cite{KoopmanLitLucas} and \cite{KoopmanLitLucasOpschoor}, which are based on dynamic copula models. The main difference between our approach and the two dynamic copula models is that we model the dependencies of latent efficient returns rather than those of observed returns. Our approach therefore does not suffer from data reduction when applied to high-frequency data. In addition, we deal explicitly with measurement errors and thus correlation estimates are not downward-biased due to microstructure effects. Multivariate GARCH generalizations have been proposed, among others, by \cite{EngleKroner}, \cite{TseTsui}, \cite{go-garch}, \cite{Carol}, \cite{EngleDCC}, \cite{GAS2}. As underlined above, asynchronicity and market microstructure effects can lead to several unwanted features in the inference of these model. We will examine in detail the impact of these two effects on the DCC model of \cite{EngleDCC} and on the $t$-GAS model of \cite{GAS2} in our simulation and empirical study. 

Score-driven models are a general class of observation-driven models where the dynamics of time-varying parameters are driven by the score of the conditional likelihood. They have been successfully applied in the recent econometric literature (see e.g. \citealt{GAS2}, \citealt{CrealSchwaabKoopmanLucas} and \citealt{OhPatton}). One of the main advantages of these models is that the conditional likelihood can be written in closed form and thus estimation can be performed through standard maximum likelihood methods. In a linear-Gaussian state-space representation, the score (with respect to system matrices) can be computed through an additional filter that runs in parallel with the Kalman filter. This method was originally introduced by \cite{GAS4} and then described in full generality by \cite{DelleMonache2}. We will use it to model the dynamics of covariances in the local-level model. The resulting model is condtionally normal and can be estimated through the Kalman filter. As pointed out by \cite{Harvey}, condtionally normal models are particularly convenient, as they feature nonlinear dynamics while preserving the possibility of applying the standard Kalman filter.  

Our Monte-Carlo analysis has three main goals. First, we investigate the finite sample properties of the maximum likelihood estimator. We find that it remains unbiased even in case many observations are missing. We then use the model as a filter for a misspecified DGP for correlations and compare it to standard dynamic models employed in a low-frequency setting. We find that, in presence of measurement errors and asynchronous observations, standard methods are subject to a downward bias. The local-level model performs significantly better, as it exploits all available data and provides robustness to measurement errors. Finally, we investigate the performance of the model in presence of fat-tails and asynchronicity and/or noise. To this end, we use the $t$-GAS model to simulate intraday prices and correlations. After randomly censoring prices, we estimate both the local-level model and the $t$-GAS.  
We find that, as the number of missing observations increases, the local-level model provides lower in-sample and out-of-sample average losses. The $t$-GAS, being a conditional correlation model for observed returns, is indeed subject to large data reduction in presence of asynchronous observations. An analogous effect is observed when we add noise to the simulated prices. Therefore, even in presence of extreme fat-tails, in an high-frequency setting the use of the local-level model is preferable to that of correctly specified observed return models.

We apply the model to transaction data of 10 NYSE stocks. The in-sample analysis based on the AIC reveals that the local-level model fits data significantly better than standard correlation models for low-frequency data. We find the well-known U-shape for volatilities, while correlations reveal an increasing pattern. The rate of increase is larger during the first two hours, then correlations increase at a slower rate and tend to decrease during the last 15 minutes of the trading day. Based on the dynamics of the first eigenvalues of the correlation matrix, we interpret this phenomenon as the emergence of a market factor which progressively explains a larger fraction of the total variance of the market.

The local-level model, being robust to asynchronous trading, can be estimated at ultra high frequencies (1-second in our application) and thus provides a description of the dynamics of covariances at very small time scales. This allows as to study the real-time response of correlations to macro-news announcements. Once macro-news arrive on the market, they are instantaneously captured by the score-driven filter, even if very few assets are traded at that time. The method can thus be employed as a nowcasting tool for high-frequency data. This interesting feature shares some similarities with the macroeconomic literature on nowcasting, where dynamic factor models are used to update forecasts of macroeconomic variables based on mixed-frequency observations (see e.g. \citealt{GIANNONE2008665} and \citealt{DelleMonache2}). In the second part of the empirical analysis, we assess the performance of the model as a nowcasting tool for high-frequency data. We construct intraday out-of-sample portfolios with short investment horizons and find that those constructed through the local-level model feature significantly lower risk.

The remaining part of the paper is organized as follows: in Section \ref{sec:model} we describe the methodology in its full generality, including the parameterization of correlations and the estimation method; Section \ref{sec:MC} discusses the results of Monte-Carlo experiments; in Section \ref{sec:empIll} we provide empirical evidence on the advantages of the model over standard techniques and study the intraday dynamic behavior of covariances; Section \ref{sec:conclusions} concludes.   

\section{Framework}
\label{sec:model}

\subsection{Model} 

Let $t\in [0,S]$ and denote by $X_t = (X_t^{(1)},\dots,X_t^{(n)})'$ an $n\times 1$ vector of intraday efficient log-prices. We consider $T$ equally-spaced observation times $0\le t_0< t_1< \dots < t_{T-1}\le S$ and propose to model $X_{t_i}$, $i=0,\dots,T-1$, as a random walk with heteroskedastic innovations:
\begin{equation}
X_{t_{i+1}} = X_{t_i} + \eta_{t_{i+1}},\quad \eta_{t_{i+1}}\sim(0,Q_{t_i})
\label{eq:transEq1}
\end{equation}
The time-varying matrix $Q_{t_i}$ describes the dynamics of volatilities and correlations of the efficient returns and is the main object of interest of this work. The efficient log-price $X_{t_i}$ is unobservable because of market microstructure effects (e.g. bid-ask bounce). Let $Y_{t_i}$ be the $n\times 1$ vector of observed log-prices. We write:
\begin{equation}
Y_{t_i}=X_{t_i}+\epsilon_{t_i},\quad \epsilon_{t_i}\sim (0,H_{t_i})
\label{eq:obsEq1}
\end{equation}
where $\epsilon_{t_i}$ is a measurement error term representing market microstructure effects. The latter is assumed to be independent from the returns of the efficient log-price. Its variances are allowed to vary over time to capture potential dynamic effects in microstructure noise. For instance, the bid-ask spread has a well-known intraday pattern, being larger at the opening hours and then declining throughout the day (\citealt{McInishWood}). Model (\ref{eq:transEq1}), (\ref{eq:obsEq1}) is at the basis of traditional market microstructure analysis of trading frictions, asymmetric information and inventory control (\citealt{Roll}, \citealt{Hasbrouck93}, \citealt{MADHAVAN2000205}).  

\subsection{State-space representation}
\label{subsec:stateSpace}
Eq. (\ref{eq:transEq1}) describes the martingale dynamics of the efficient log-price process, while eq. (\ref{eq:obsEq1}) is the associated measurement equation. We can re-write both equations as:
\begin{align}
Y_{t} &= X_{t}+ \epsilon_{t},\quad \epsilon_{t}\sim (0,H_{t}) \label{eq:ssm:1}\\
X_{t+1} &= X_{t} + \eta_{t},\quad \eta_{t}\sim (0,Q_{t}) \label{eq:ssm:2}
\end{align}  
where, without loss of generality, we have set $t_{i+1}-t_i=1$. 
Model (\ref{eq:ssm:1}), (\ref{eq:ssm:2}) is known as \textit{local-level} model (\citealt{Harvey},\citealt{DurbinKoopman}). If the two covariance matrices $H_t$ and $Q_t$ are constant, the local-level model can be estimated by quasi-maximum likelihood through the Kalman filter. This is the route of \cite{KEM} and \cite{ShephardXiu}, who proposed a quasi-maximum likelihood estimator of the integrated covariance of high-frequency asset prices. Here we are interested in a different problem, namely the dynamic modelling of the covariances of both the noise and the efficient returns. We therefore need to specify a dynamic equation for $H_t$ and $Q_t$.


A convenient way to handle model (\ref{eq:ssm:1}), (\ref{eq:ssm:2}) is assuming that the disturbance terms are condtionally normal (\citealt{Harvey}). Let $\mathcal{F}_{t-1}$ be the $\sigma$-field generated by observations of the log-price process up to time $t-1$. The condtionally normal local-level model reads:
\begin{align}
 Y_{t} &=X_{t}+ \epsilon_{t},\quad \epsilon_{t}|\mathcal{F}_{t-1}\sim\text{NID} (0,H_{t}) \label{eq:ll:1}\\
X_{t+1} &= X_{t} + \eta_{t},\quad \eta_{t}|\mathcal{F}_{t-1}\sim\text{NID}(0,Q_{t}) \label{eq:ll:2}
\end{align}
meaning that, conditionally on the information available at time $t-1$, the distribution of $\epsilon_t$ and $\eta_t$ is normal, with known covariance matrices $H_t$ and $Q_t$. The latter are assumed to obey an observation-driven update rule and depend nonlinearly on past observations. The Kalman filter can be applied to compute the likelihood in the usual prediction error form. As discussed by \cite{Harvey}, condtionally normal models allow to ``inject'' nonlinear dynamics in the model while still preserving the possibility of applying the standard Kalman filter.
In our empirical framework, this is a substantial advantage, given that the Kalman filter can easily handle missing values and can reconstruct the efficient price based on all available observations.    

\subsection{Time-varying covariances}
In order to estimate the model, we need to specify the law of motion of $H_t$ and $Q_t$. Score-driven models (\citealt{GAS1}, \citealt{Harvey_2013}) are a general class of observation-driven models. In score-driven models, parameters are updated based on the score of the conditional density. The GARCH model of \cite{Bollerslev}, the EGARCH model of \cite{Nelson} and the ACD model of \cite{EngleRussel} are examples of models that can be recovered in this general framework. In addition, score-driven models have information theoretic optimality properties, as shown by \cite{Blasques}. 

Motivated by the flexibility of score-driven models, we choose to model the dynamics of $H_t$ and $Q_t$ based on the score of the conditional density. The covariance matrix $Q_t$ can be decomposed as:
\begin{equation}
 Q_t=D_tR_t D_t
\end{equation}
where $D_t=\text{diag}[Q_t]^{1/2}$ is a diagonal matrix of standard deviations and $R_t$ is a correlation matrix. This decomposition is common in the econometric literature and is used, for instance, in the DCC model of \cite{EngleDCC}. For parsimony, we assume that the covariance matrix $H_t$ of the noise is diagonal. This assumption can be relaxed at the expense of increasing considerably the number of time-varying parameters. However, as pointed out by \cite{KEM}, in the static case the off-diagonal elements of $H_t$ are found to be close to zero. They thus assumed a diagonal covariance matrix for the noise. \cite{ShephardXiu} made the same assumption.

Let $f_t$ denote a vector of time-varying parameters. We write:
\begin{equation}
 f_t = 
 \begin{pmatrix}
  \log(\text{diag}[H_t]) \\
  \log(\text{diag}[D_t^2])\\
  \phi_t
 \end{pmatrix}
 \label{eq:ft}
\end{equation}
where $\phi_t$ is a $q\times 1$ vector depending on the parameterization of the correlation matrix $R_t$. The latter will be discussed in Section \ref{subsec:parameterization}. The number of components of $f_t$ is thus $k=2n+q$. The update rule in score-driven models is given by:
\begin{equation}
 f_{t+1}=\omega + As_t + Bf_t
 \label{eq:gas}
\end{equation}
where $s_t$ is the scaled score vector:
\begin{equation}
s_t = (\mathcal{I}_{t|t-1})^{-1}\nabla_t,\quad \nabla_t=\left[\frac{\partial{\log p(Y_t|f_{t},\mathcal{F}_{t-1},\Theta)}}{\partial{f_t'}}\right]',\quad \mathcal{I}_{t|t-1}=\text{E}[\nabla_t\nabla_t']
\end{equation}
The $k\times 1$ vector $\omega$ and the $k\times k$ matrices $A,B$ are included in the vector $\Theta$ of static parameters to be estimated. 

The conditional log-likelihood is given by:
\begin{equation}
 \log p(Y_t|f_{t},\mathcal{F}_{t-1},\Theta) = \text{const}-\frac{1}{2}\left(\log|F_t|+v_t'F_t^{-1}v_t\right)
 \label{eq:condLogLik}
\end{equation}
where $v_t$ and $F_t$ are the Kalman filter prediction error and its covariance matrix, as defined in Appendix \ref{app:kf}. As shown by \cite{DelleMonache2}, the score $\nabla_t$ and the Fisher information matrix $\mathcal{I}_{t|t-1}$ can be computed as:
\begin{align}
\nabla_t &= -\frac{1}{2}\left[\dot{F}_t'(\mathcal{I}_{n_t}\otimes F_t^{-1})\text{vec}(\mathcal{I}_{n_t}-v_tv_t'F_t^{-1})+2\dot{v}_t'F_t^{-1}v_t\right]\label{eq:nabla}\\
\mathcal{I}_{t|t-1} &= \frac{1}{2}\left[\dot{F}_t'(F_t^{-1}\otimes F_t^{-1})\dot{F}_t + 2\dot{v}_t'F_t^{-1}\dot{v_t}\right]\label{eq:fisher}
\end{align}
where $n_t$ denotes the number of observations available at time $t$. Together with $v_t$ and $F_t$, the computation of $\nabla_t$ and $\mathcal{I}_{t|t-1}$ requires $\dot{v}_t$ and $\dot{F}_t$, which denote derivatives of $v_t$ and $F_t$ with respect to the time-varying parameter vector $f_t$. As discussed by \cite{DelleMonache2}, they can be computed through the recursions reported in Appendix \ref{app:kf}. 

By running in parallel the Kalman filter and the filter in eq. (\ref{eq:gas}), one can update parameters and compute the conditional log-likelihood in eq. (\ref{eq:condLogLik}). The static parameters $\Theta$ are estimated by numerically optimizing the log-likelihood function:
\begin{equation}
 \hat{\Theta} = \argmax\limits_{\Theta}\sum_{t=1}^T \log p(Y_t|f_{t},\mathcal{F}_{t-1},\Theta) 
\end{equation}
Restrictions on the structure of $A$, $B$ are discussed in the empirical application in Section \ref{sec:empIll}.

\subsection{Parameterization of \texorpdfstring{$R_t$}{Lg}}
\label{subsec:parameterization}
To have a full model specification, we need a parameterization for the correlation matrix $R_t$. We restrict our attention to parameterizations that guarantee a positive-definite correlation matrix. One possibility is to use hyperspherical coordinates, as in \cite{GAS2}. Another possibility is given by the equicorrelation parameterization of \cite{EngleKelly}.

In the first case, we write the correlation matrix as $R_t = Z_t'Z_t$. The matrix $Z_t$ has the form:
\begin{equation}
Z_t =
\begin{pmatrix}
1  &  c_{12}  &      c_{13}      & \dots &  c_{1n} \\
0  &  s_{12}  &  c_{23}s_{13}    & \dots &  c_{2n}s_{1n} \\
0  &  0       &  s_{23}s_{13}    & \dots &  c_{3n}s_{2n}s_{1n} \\
\vdots & \vdots & \vdots & & \vdots \\
0 & 0 & 0 & \dots & \prod_{k=1}^{n-1}s_{kn} 
\end{pmatrix}
\end{equation}
where $c_{ij}=\cos{\theta_{ij}}$ and $s_{ij} = \sin{\theta_{ij}}$. Note that the time index has been omitted for ease of notation. The $i$-th column of $Z_t$ contains the hyperspherical coordinates of a vector of unit norm in an $i$-th dimensional subspace of $\mathbb{R}^n$, which is parametrized by $i-1$ angles. We have therefore $n(n-1)/2$ angles, equal to the number of correlations in $R_t$. We set $\phi_t=[\theta_{12,t},\theta_{13,t},\dots,\theta_{n-1n,t}]'$ in eq. (\ref{eq:ft}), so that $q=n(n-1)/2$. The number of time-varying parameters is thus $k=2n+n(n-1)/2$. 

In the equicorrelation parameterization, the correlation matrix $R_t$ is written as:
\begin{equation}
 R_t = (1-\rho_t)\mathcal{I}_n + \rho_t\mathcal{J}_n
\end{equation}
where $\mathcal{J}_n$ denotes an $n\times n$ matrix of ones. The correlation matrix $R_t$ is positive-definite if and only if the parameter $\rho_t$ satisfies the constraint $-1/(n-1)\le \rho_t\le 1$. One possibility to guarantee this constraint is to write:
\begin{equation}
 \rho_t = \frac{1}{2}\left[\left(1-\frac{1}{n-1}\right)+\left(1+\frac{1}{n-1}\right)\text{tanh}(\theta_t)\right]
\end{equation}
as in \cite{KoopmanLitLucasOpschoor}. We set $\phi_t=\theta_t$, so that $q=1$. The number of time-varying parameters is thus $k=2n+1$. The equicorrelation parameterization is very parsimonious, as it assumes the same correlation for all couples of assets. Notwithstanding, it has been proven to be effective in several empirical problems (cf. discussions in \citealt{EngleKelly}).

\section{Monte-Carlo analysis}
\label{sec:MC}

\subsection{Finite-sample properties}
\label{sub:MC:finiteSample}

We first study the finite sample properties of the maximum likelihood estimator through Monte-Carlo simulations. We set $n=10$, the same number of assets that will be used in our empirical application in Section \ref{sec:empIll}. The number of time-varying parameters is thus $k=2n+n(n-1)/2=65$ in the parameterization with hyperspherical coordinates and $k=2n+1=21$ in the equicorrelation parameterization. In the first experiment, we assume that the static parameters $\omega$, $A$, $B$ have the following structure:
\begin{equation}
 \omega = 
 \begin{pmatrix}
  \omega^{\text{h}}\\
  \omega^{\text{d}}\\
  \omega^{\text{r}}\\
 \end{pmatrix}
 ,\quad
 A =\text{diag}
 \begin{pmatrix}
  A^{\text{h}} \\
  A^{\text{d}} \\
  A^{\text{r}} \\
 \end{pmatrix}
,\quad
 B = \text{diag}
 \begin{pmatrix}
  B^{\text{h}} \\
  B^{\text{d}} \\
  B^{\text{r}}\\
 \end{pmatrix}
 \label{eq:ThetaUnrestr}
 \end{equation}
where $\omega^{\text{h}}$, $A^{\text{h}}$, $B^{\text{h}}$ are $n\times 1$ vectors driving the variances of the noise, $\omega^{\text{d}}$, $A^{\text{d}}$, $B^{\text{d}}$ are $n\times 1$ vectors driving the variances of efficient returns and $\omega^{\text{r}}$, $A^{\text{r}}$, $B^{\text{r}}$ are $q\times 1$ vectors driving the correlations. We further constraint the elements in each vector to be equal among each other. The number of static parameters is therefore equal to $9$. In particular, we set: $\omega^{\text{h}}=-0.0461\iota_n$, $\omega^{\text{d}}=-0.0322\iota_n$, $\omega^{\text{r}}= 0.0185\iota_q$, $A^{\text{h}}=A^{\text{d}} = 0.02\iota_n$, $A^{\text{r}} = 0.02\iota_q$, $B^{\text{h}}=B^{\text{d}}= 0.98\iota_n$, $B^{\text{r}}=0.98\iota_q$. The values of $\omega^{\text{h}}$, $\omega^{\text{d}}$ are chosen in such a way that the signal-to-noise ratio (defined as the ratio between the unconditional variance of the efficient returns and the unconditional variance of the measurement error) is similar to the one found on empirical data, which is on average equal to $1$ (cf. table \ref{tab:dataset}).

After simulating the log-prices, we randomly censor observations to mimic asynchronous trading. The probability of removing one observation is denoted by $\lambda$ and is assumed to be the same for the $n$ time-series. To set the initial values of time-varying parameters, we estimate a local-level model with constant parameters in the subsample comprising the first 100 observations. This can be done through the EM algorithm, as described by \cite{KEM}. We then choose $f_1$ based on eq. (\ref{eq:ft}), with $H_t$, $D_t$, $R_t$ equal to those obtained through the EM algorithm. We simulate $N=1000$ time-series of $T=2000$ observations and consider for scenarios characterized by $\lambda=0$, $0.3$, $0.5$, $0.8$. Table \ref{tab:simHyper} reports summary statistics of maximum likelihood estimates in the parameterization with hyperspherical coordinates. Table \ref{tab:simEqui} reports analogous statistics obtained with the equicorrelation parameterization. 
\bigskip 
\begin{table}[!htbp]
\centering
\small
\setlength{\tabcolsep}{12pt}
\begin{tabular}{c|cccccc}
  \hline
  $\lambda$ & Mean & Std & Mean & Std & Mean & Std  \\
  \hline
   & \multicolumn{2}{c}{$\omega^{\text{h}}=-0.0461\iota_n$} & \multicolumn{2}{c}{$\omega^{\text{d}}=-0.0322\iota_n$} & \multicolumn{2}{c}{$\omega^{\text{r}}=0.0185\iota_q$} \\
   0.0 &  -0.0460  &  0.0005 & -0.0322 &  0.0012 & 0.0186 &  0.0004 \\
   0.3 & -0.0482 & 0.0140 & -0.0330 & 0.0073 & 0.0200 & 0.0029 \\
   0.5 & -0.0497 & 0.0161 & -0.0349 & 0.0117 & 0.0204 &  0.0039\\
   0.8 & -0.0510 & 0.0170 & -0.0351 & 0.0122 & 0.0209 & 0.0045 \\
   \hline
   & \multicolumn{2}{c}{$A^{\text{h}}=0.02\iota_n$} & \multicolumn{2}{c}{$A^{\text{d}}=0.02\iota_n$} & \multicolumn{2}{c}{$A^{\text{r}}=0.02\iota_q$} \\
   0.0 & 0.0200 & 0.0001 & 0.0200 & 0.0004 & 0.0200 & 0.0002 \\
   0.3 & 0.0214 & 0.0047 & 0.0200 & 0.0031 & 0.0193 & 0.0016\\
   0.5 & 0.0231 & 0.0075 & 0.0201 & 0.0039 & 0.0203 & 0.0021 \\
   0.8 & 0.0234 & 0.0080 & 0.0205 & 0.0042 & 0.0211 & 0.0032 \\
     \hline
   & \multicolumn{2}{c}{$B^{\text{h}}=0.98\iota_n$} & \multicolumn{2}{c}{$B^{\text{d}}=0.98\iota_n$} & \multicolumn{2}{c}{$B^{\text{r}}=0.98\iota_q$} \\
   0.0 & 0.9800 & 0.0002 & 0.9799 & 0.0008 & 0.9800 & 0.0003 \\
   0.3 & 0.9790 & 0.0061 & 0.9797 & 0.0045 & 0.9790 & 0.0031 \\
   0.5 & 0.9784 & 0.0070 & 0.9785 & 0.0072 & 0.9788 & 0.0040\\
   0.8 & 0.9780 & 0.0075 & 0.9779 & 0.0080 & 0.9785 & 0.0045 \\
   \hline  
\end{tabular}
\caption{Mean and standard deviations of maximum likelihood estimates of the local-level model with score-driven covariances in the parameterization with hyperspherical coordinates.}
\label{tab:simHyper}
\end{table}

The results show that, in both parameterizations, maximum likelihood estimates concentrate around the true parameters. Not surprisingly, missing values lead to an increase of the variance of the maximum likelihood estimator. However, even in the highly asynchronous scenario with $\lambda=0.8$, relative errors remain small for all parameters in $A$, $B$ and for almost all parameters in the intercept $\omega$. Larger relative errors are found for the parameter $\omega^{\text{r}}$ in the equicorrelation parameterization. 
\bigskip 
\begin{table}[!htbp]
\centering
\small
\setlength{\tabcolsep}{12pt}
\begin{tabular}{c|cccccc}
  \hline
  $\lambda$ & Mean & Std & Mean & Std & Mean & Std  \\
  \hline
   & \multicolumn{2}{c}{$\omega^{\text{h}}=-0.0461\iota_n$} & \multicolumn{2}{c}{$\omega^{\text{d}}=-0.0322\iota_n$} & \multicolumn{2}{c}{$\omega^{\text{r}}=0.0185\iota_q$} \\
   0.0 & -0.0489 & 0.0111 & -0.0338 & 0.0096 & 0.0201 & 0.0098\\
   0.3 & -0.0507 & 0.0128 & -0.0339 & 0.0121 & 0.0209 & 0.0111 \\
   0.5 & -0.0525 & 0.0137 & -0.0402 & 0.0188 & 0.0211 & 0.0130\\
   0.8 & -0.0560 & 0.0145 & -0.0422 & 0.0198 & 0.0215 & 0.0134 \\
   \hline
   & \multicolumn{2}{c}{$A^{\text{h}}=0.02\iota_n$} & \multicolumn{2}{c}{$A^{\text{d}}=0.02\iota_n$} & \multicolumn{2}{c}{$A^{\text{r}}=0.02\iota_q$} \\
   0.0 & 0.0200 & 0.0043 & 0.0205 & 0.0034 & 0.0205 & 0.0054 \\
   0.3 & 0.0186 & 0.0049 & 0.0184 & 0.0037 & 0.0198 & 0.0053\\
   0.5 & 0.0171 & 0.0060 & 0.0185 & 0.0042 & 0.0196 & 0.0065 \\
   0.8 & 0.0170 & 0.0080 & 0.0178 & 0.0050 & 0.0194 & 0.0070 \\
     \hline
   & \multicolumn{2}{c}{$B^{\text{h}}=0.98\iota_n$} & \multicolumn{2}{c}{$B^{\text{d}}=0.98\iota_n$} & \multicolumn{2}{c}{$B^{\text{r}}=0.98\iota_q$} \\
   0.0 & 0.9788 & 0.0070 & 0.9790 & 0.0059 & 0.9746 & 0.0142 \\
   0.3 & 0.9780 & 0.0077 & 0.9789 & 0.0074 & 0.9749 & 0.0140 \\
   0.5 & 0.9772 & 0.0090 & 0.9749 & 0.0118 & 0.9719 & 0.0191\\
   0.8 & 0.9770 & 0.0110 & 0.9732 & 0.0132 & 0.9702 & 0.0201 \\
   \hline  
\end{tabular}
\caption{Mean and standard deviations of maximum likelihood estimates of the local-level model with score-driven covariances in the parameterization based on the equicorrelation matrix.}
\label{tab:simEqui}
\end{table}
We then assume parameters $\omega$, $A$, $B$ have the following structure:
\begin{equation}
 \omega = 
 \begin{pmatrix}
  0\\
  \vdots\\
  0\\
 \end{pmatrix}
 ,\quad
 A =\text{diag}
 \begin{pmatrix}
  A^{\text{h}} \\
  A^{\text{d}} \\
  A^{\text{r}} \\
 \end{pmatrix}
,\quad
 B = \text{diag}
 \begin{pmatrix}
  1 \\
  \vdots \\
  1\\
 \end{pmatrix}
  \label{eq:ThetaRW}
 \end{equation}
Under these restrictions, time-varying parameters have non-stationary random-walk dynamics. As will be shown in Section \ref{sub:structStatic}, this specification provides a better description of intraday high-frequency prices and will be used to estimate the local-level model on empirical data. We set $A^{\text{h}}=A^{\text{d}} = 0.02\iota_n$, $A^{\text{r}} = 0.02\iota_q$ and report in Table \ref{tab:simRw} summary statistics of maximum likelihood estimates. As in the previous simulation, parameter estimates concentrate around true values and the variance slightly increases with the number of missing values. Overall, maximum likelihood provides satisfactory parameter estimates in all the considered scenarios.

\bigskip 
\begin{table}[!htbp]
\centering
\small
\setlength{\tabcolsep}{12pt}
\begin{tabular}{c|cccccc}
  \hline
  $\lambda$ & Mean & Std & Mean & Std & Mean & Std  \\
  \hline
  & \multicolumn{6}{c}{\underline{\textsc{hyperspherical}}}\\
   & \multicolumn{2}{c}{$A^{\text{h}}=0.02\iota_n$} & \multicolumn{2}{c}{$A^{\text{d}}=0.02\iota_n$} & \multicolumn{2}{c}{$A^{\text{r}}=0.02\iota_q$} \\
   0.0 & 0.0200 & 0.0003 & 0.0200 & 0.0004 & 0.0203 & 0.0008 \\
   0.3 & 0.0201 & 0.0019 & 0.0194 & 0.0012 & 0.0199 & 0.0010\\
   0.5 & 0.0204 & 0.0030 & 0.0190 & 0.0022 & 0.0197 & 0.0011 \\
   0.8 & 0.0202 & 0.0030 & 0.0192 & 0.0029 & 0.0199 & 0.0015 \\
     \hline
      & \multicolumn{6}{c}{\underline{\textsc{equicorrelation}}}\\
   & \multicolumn{2}{c}{$A^{\text{h}}=0.02\iota_n$} & \multicolumn{2}{c}{$A^{\text{d}}=0.02\iota_n$} & \multicolumn{2}{c}{$A^{\text{r}}=0.02\iota_q$} \\
   0.0 & 0.0200 & 0.0005 & 0.0203 & 0.0006 & 0.0200 & 0.0003 \\
   0.3 & 0.0191 & 0.0020 & 0.0198 & 0.0022 & 0.0199 & 0.0012\\
   0.5 & 0.0181 & 0.0039 & 0.0189 & 0.0032 & 0.0196 & 0.0022 \\
   0.8 & 0.0179 & 0.0045 & 0.0185 & 0.0038 & 0.0196 & 0.0025 \\
   \hline  
\end{tabular}
\caption{Mean and standard deviations of maximum likelihood estimates of the local-level model with score-driven covariances and random walk-type restrictions for static parameters.}
\label{tab:simRw}
\end{table}

\subsection{Assessing the effect of noise and asynchronicity}
\label{sub:MC:noiseAyn}
In this Section we aim to assess the effect of noise and asynchronous observations on commonly used dynamic correlation models and to show the advantages provided by the proposed modelling strategy. High-frequency prices are typically synchronized before being analyzed with standard dynamic models. The most popular synchronization scheme is previous-tick interpolation, consisting in updating missing values with previous available prices. This procedure (and other similar schemes) leads to a downward bias of correlations toward zero. In the literature on realized covariance estimation, the latter is known as ``Epps effect'' (\citealt{Epps}). The explanation of the Epps effect is intuitive: synchronization leads to a large number of zero returns, which in turn undermine correlations (see \citealt{HayashiYoshida} and references therein). As will be shown here, a similar problem arises when estimating dynamic correlations. The presence of measurement errors is an additional source of bias for correlations. This is not surprising, as the attenuation bias due to measurement errors occurs in several econometrics and statistics problems, e.g. in \textit{error-in-variables} models. 

\begin{figure}
\centering
    \includegraphics[width=1\linewidth]{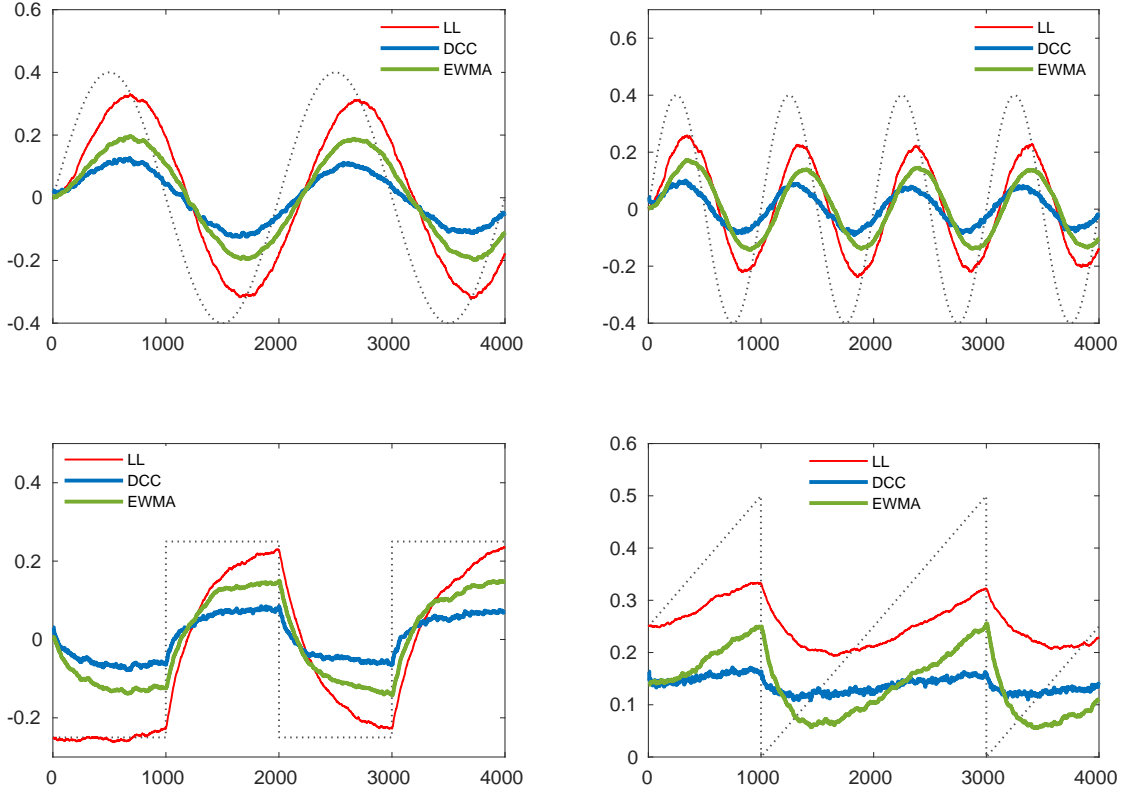}
    \caption{Estimates of local-level, DCC and EWMA averaged over all simulations in the scenario $\delta=2$, $\lambda=0.5$.}
    \label{fig:simIncAsyn}
\end{figure}

For simplicity, we consider a bivariate model generated as follows:
\begin{align}
Y_t & = X_t  + \epsilon_t,\quad \epsilon_t \sim\textit{N}(0,H)\\
 X_{t+1} &= X_t + \eta_t,\quad \eta_t \sim\textit{N}(0,D R_tD)
\end{align}
where $H = h\mathcal{I}_2$, $D = d\mathcal{I}_2$ and:
\begin{equation}
R_t = 
 \begin{pmatrix}
 1 & \rho_t\\
 \rho_t & 1
 \end{pmatrix}
\end{equation}
The correlation coefficient $\rho_t$ follows these dynamic patterns:
\begin{enumerate}
\item Sine \hspace{1.5cm} $\rho_t =  b_s\sin(c_s\pi t)$
\item Fast Sine \hspace{0.7cm} $\rho_t = b_f\sin(c_f\pi t)$
\item Step \hspace{1.5cm} $\rho_t = \alpha_s - \beta_s(\mathbbm{1}_{t<T/4}+\mathbbm{1}_{T/2<t<3T/4})$
\item Ramp \hspace{1.2cm} $\rho_t = \frac{1}{b_r}\mod(t+a_r,c_r)$
\item Model \hspace{1.2cm} $\rho_t = \exp{(h_t)}/[1+\exp{(h_t)}]$
\end{enumerate}
where $h_t$ is an AR$(1)$ process:
\begin{equation*}
h_{t+1} = c_m + b_mh_t + a_m\phi_t,\quad \phi_t\sim \text{N}(0,1)
\end{equation*}
and $T$ is the number of observations. The parameters governing the dynamics of $\rho_t$ are chosen as: $b_s=b_f=0.4$, $c_s = 4/T$, $c_f = 8/T$, $\alpha_s=0.25$, $\beta_s = 0.5$, $b_r = T$, $a_r = b_r/4$, $c_r=b_r/2$, $b_m = 0.99$, $c_m = -0.4(1-b_m)$, $a_m = 0.01$. The variance $d^2$ of the latent process is constant and equal to $0.1$ for all the simulated patterns. The variance $h$ of the noise is computed based on the signal-to-noise ratio $\delta=d^2/h$. We consider three different scenarios with $\delta = 0.5$ (low signal), $1$ (moderate signal), $2$ (high signal). As will be evident in Section \ref{sec:empIll}, these values are close to those estimated on real data. Once observations have been generated, we apply a censoring scheme similar to the one used in the previous analysis. We consider three scenarios with $\lambda = 0$, $0.3$, $0.5$, where $\lambda$ denotes the probability of removing an observation. 

We generate $N=250$ simulations of $T=4000$ observations for each dynamic pattern and for each scenario. As a benchmark, we consider the DCC model of \cite{EngleDCC} and an EWMA scheme given by:
\begin{equation}
 \hat{Q}_{t+1} = \gamma \hat{Q}_t + (1-\gamma)r_t r_t'
\end{equation}
where $\gamma$ is set equal to $0.96$. Both DCC and EWMA are applied after synchronizing data through previous-tick interpolation. Depending on the value of $\lambda$, this results in a large number of zero returns. In order to attenuate the effect of noise and zero returns on these two models, we sample observations at a lower frequency. This procedure is commonly employed when computing e.g. realized covariance. Indeed, at lower frequencies, returns are less affected by both microstructure effects and by staleness due to absence of trading. A natural consequence is that a significant part of data is neglected. However, subsampling greatly improves correlation estimates of DCC and EWMA. In each scenario, the new sampling frequency is chosen as the one minimizing the MSE in a pre-simulation study with $N=20$ simulations. The selected values range in the interval between $3$ and $6$ time units. 

\begin{figure}
\centering
    \includegraphics[width=1\linewidth]{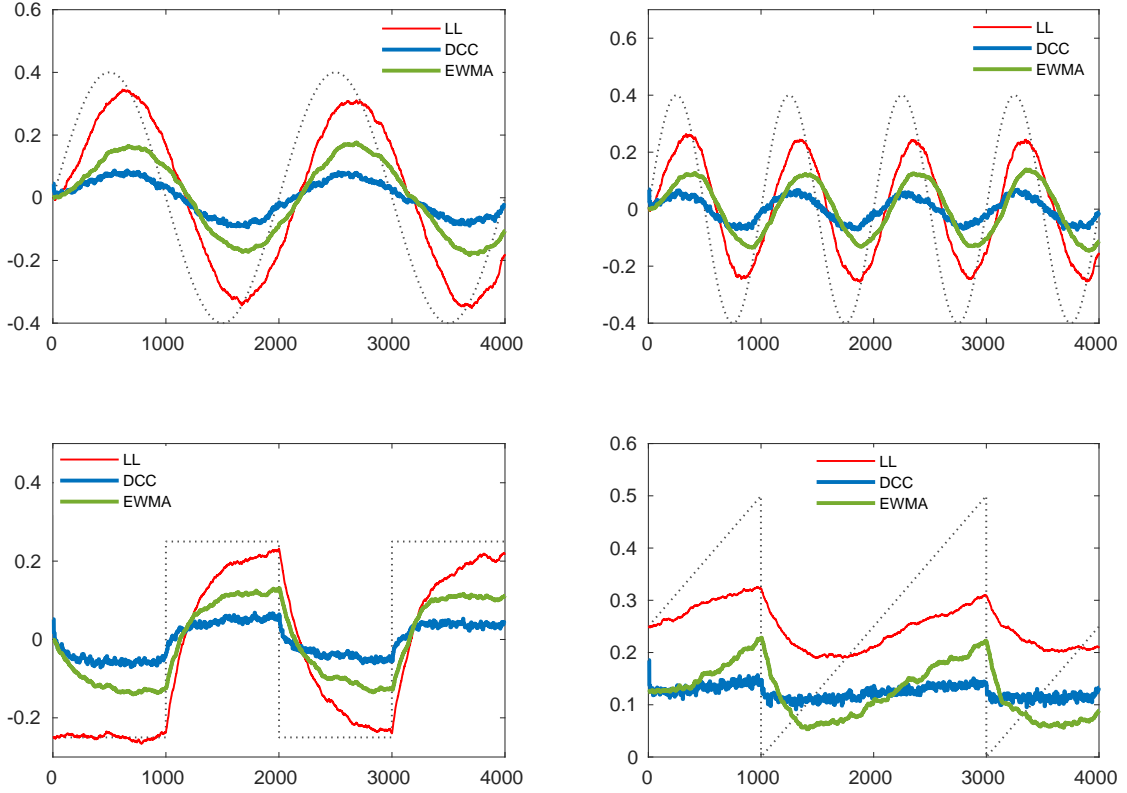}
    \caption{Estimates of local-level, DCC and EWMA averaged over all simulations in the scenario $\delta=0.5$, $\lambda=0$.}
    \label{fig:simIncNoise}
\end{figure}

We compute both in-sample and out-of-sample losses of the local-level model and the DCC. To this end, the sample is divided into two subsamples containing the same number of observations. The latter is equal to $T_{\text{sub}}=2000$ in the local-level model, while it depends on the sampling frequency in the case of the DCC. The local-level model and the DCC are estimated in the first subsample, where in-sample losses are computed. The second subsample is then used to compute out-of-sample losses. The parameterization used in the local-level model is the one based on hyperspherical coordinates. Figure \ref{fig:simIncAsyn} shows, in the scenario $\delta = 2, \lambda = 0.5$, correlation estimates of the three models averaged over all simulations. Even after sampling at lower frequencies, DCC and EWMA estimates are biased toward zero. This is mainly a consequence of asynchronicity, which is high in this scenario. Figure \ref{fig:simIncNoise} shows, in the scenario $\delta=0.5$, $\lambda=0$, average estimates of the three models. DCC and EWMA have a similar downward bias. Compared to the previous scenario, the bias is now entirely due to measurement errors, as asynchronicity is absent. It is instead evident that the local-level model is much less affected by the two effects. The use of standard dynamic covariance models thus leads to data reduction, as a consequence of sampling at lower frequencies, and underestimates (absolute value of) correlations.

Table \ref{tab:simIncNoiseAsyn} reports in-sample and out-of-sample MSE of the three models in all the simulated scenarios. The local-level significantly outperforms the DCC in all the scenarios. Compared to the DCC, the performance of the EWMA is closer to that of the local-level in patterns ``sine'', ``fast sine'', ``step''. However, with only one exception (out-of-sample MSE of ``fast sine'' in scenario $\lambda=0.5$, $\delta=2$) the MSE of the local-level is lower than that of the EWMA. For patterns ``ramp'' and ``model'', characterized by frequent changes and stochasticity, the local-level performs substantially better than the EWMA.

\begin{sidewaystable}
\centering
\footnotesize
\begin{tabular}{c|lllll|lllll|lllll} 
Model &  Sine  & Fast  & Step & Ramp & Model & Sine  & Fast & Step & Ramp & Model & Sine  & Fast  & Step & Ramp & Model\\ \hline
&  \multicolumn{5}{|c|}{$\delta = 0.5$} & \multicolumn{5}{c|}{$\delta = 1$}  & \multicolumn{5}{c}{$\delta = 2$}  \\
\hline
& \multicolumn{15}{c}{$\lambda = 0$}\\
\hline
LL & 0.0367 & 0.0586 & 0.0302 & 0.0217  & 0.0361 & 0.0263  & 0.0413 & 0.0234 & 0.0200 & 0.0279 & 0.0197 & 0.0302 & 0.0185 & 0.0178 & 0.0224\\ 
DCC & 0.0628 & 0.0711 & 0.0533 & 0.0422  & 0.1222 & 0.0455 & 0.0575 & 0.0409 & 0.0295 & 0.0760 & 0.0344 & 0.0457 & 0.0327 & 0.0225 & 0.0474\\ 
EWMA & 0.0474 & 0.0643 & 0.0388 & 0.0402 & 0.1066 & 0.0376 & 0.0586 & 0.0317 & 0.0290 & 0.0641 & 0.0321 & 0.0555 & 0.0277 & 0.0230 & 0.0417\\ 
\hline
LL & 0.0438 & 0.0620 & 0.0435 & 0.0237 & 0.0458 &  0.0302 & 0.0453  & 0.0370 & 0.0216 & 0.0309 & 0.0219 & 0.0325 & 0.0290 & 0.0187 & 0.0226      \\ 
DCC & 0.0655 & 0.0718 & 0.0570 & 0.0421 & 0.1283 & 0.0471 & 0.0582  & 0.0455 & 0.0296 & 0.0811 & 0.0352 & 0.0459 & 0.0376 & 0.0228 & 0.0516 \\ 
EWMA & 0.0489 & 0.0677 & 0.0462 & 0.0399 & 0.1068 & 0.0405 & 0.0631 & 0.0402 & 0.0291 & 0.0630 & 0.0362 & 0.0609 & 0.0370 & 0.0235 & 0.0400\\
\hline
& \multicolumn{15}{c}{$\lambda = 0.3$}\\
\hline
LL & 0.0482 & 0.0673 & 0.0354 & 0.0228 & 0.0456 & 0.0359 & 0.0551 & 0.0280  & 0.0219 & 0.0352 & 0.0274 & 0.0426 & 0.0225 & 0.0203 & 0.0293 \\
DCC & 0.0664 & 0.0760 & 0.0581 & 0.0465 & 0.1192 & 0.0524 & 0.0645 & 0.0476 & 0.0343 & 0.0903 & 0.0429 & 0.0548 & 0.0407 & 0.0272 & 0.0646 \\
EWMA & 0.0506 & 0.0771 & 0.0432 & 0.0450 & 0.1103 & 0.0419  & 0.0615 & 0.0363 & 0.0335 & 0.0829 & 0.0365  & 0.0582 & 0.0319 & 0.0266 & 0.0603 \\
\hline
LL & 0.0498 & 0.0701 & 0.0584 & 0.0242 & 0.0534 & 0.0414 & 0.0618 & 0.0443 & 0.0222 & 0.0405 & 0.0308 & 0.0472 & 0.0392 & 0.0202 &  0.0322 \\
DCC & 0.0704 & 0.0775 & 0.0617 & 0.0464 & 0.1441 & 0.0559 & 0.0661 & 0.0522 & 0.0345 & 0.1153 & 0.0450 & 0.0564 & 0.0453 & 0.0277 & 0.0839 \\
EWMA & 0.0533 & 0.0792 & 0.0593 & 0.0452 & 0.1261 & 0.0443 & 0.0656 & 0.0479 & 0.0345 & 0.0938 & 0.0391 & 0.0630 & 0.0410 & 0.0280 & 0.0653 \\
\hline
& \multicolumn{15}{c}{$\lambda = 0.5$}\\
\hline 
LL & 0.0540 & 0.0676 & 0.0446 & 0.0247 & 0.0489 & 0.0452 & 0.0622 & 0.0346 & 0.0229 & 0.0407 &  0.0362 & 0.0600 & 0.0281 & 0.0214 & 0.0354\\
DCC & 0.0692 & 0.0760 & 0.0600 & 0.0513 & 0.1465 & 0.0595 & 0.0683 & 0.0527 & 0.0405 & 0.1124 & 0.0512 & 0.0615 & 0.0469 & 0.0338 & 0.0887 \\
EWMA & 0.0560 & 0.0699 & 0.0469 & 0.0510 & 0.1402 & 0.0463  & 0.0645 & 0.0406 & 0.0405 & 0.1058 & 0.0413 & 0.0611 & 0.0364 & 0.0337 & 0.0834 \\
\hline
LL & 0.0684 & 0.0690 & 0.0525 & 0.0261 & 0.0611 & 0.0548 & 0.0692 & 0.0464 &  0.0240 & 0.0505 & 0.0435 & 0.0680 & 0.0417 & 0.0226 & 0.0424 \\
DCC & 0.0738 & 0.0780 & 0.0623 & 0.0519 & 0.1844 & 0.0644 & 0.0709 & 0.0560 & 0.0412 & 0.1438 & 0.0555 & 0.0643 & 0.0509 & 0.0346 & 0.1149\\
EWMA & 0.0589 & 0.0737 & 0.0537 & 0.0529 & 0.1635 & 0.0575 & 0.0691 & 0.0479 & 0.0423 & 0.1205 & 0.0450 & 0.0662 & 0.0442 & 0.0354 & 0.0924 \\
\hline
\end{tabular}
\renewcommand{\baselinestretch}{1.2}
\caption{MSE of local-level (LL), DCC and EWMA estimates in the 9 scenarios obtained by combining the three missing value scenarios ($\lambda =0, 0.3,0.5$) to each of the three signal scenarios ($\delta=0.5,1,2$). The first three lines of each scenario show in-sample results, while the last three lines report out-of-sample results.}
\label{tab:simIncNoiseAsyn}  
\end{sidewaystable} 
\restoregeometry

\subsection{Comparison with fat-tail return models}
\label{sub:MC:outliers}

High-frequency data can exhibit extreme movements during flash crashes or in correspondence of macro-news announcements. It is therefore important to investigate the quality of correlation estimates provided by the proposed approach in presence of outliers. \cite{GAS2} introduced a score-driven dynamic correlation model, named $t$-GAS, based on multivariate Student-$t$ distribution. The conditional density of log-returns in the $t$-GAS is:

\begin{equation}
 p(r_t|\Sigma_t,\nu) = \frac{\Gamma(\frac{\nu+n}{2})}{\Gamma(\frac{\nu}{2})[(\nu-2)\pi]^{n/2}|\Sigma_t|^{1/2}}\left(1+\frac{r_t'\Sigma_t r_t}{\nu-2}\right)^{-(\nu+n)/2} 
 \label{eq:tGAS}
\end{equation}
The covariances in $\Sigma_t$ obey the usual score-driven update rule. Different parameterizations for $\Sigma_t$ are possible, such as the one discussed in Section \ref{subsec:parameterization} based on hyperspherical coordinates. 

In this simulation study, we use eq. (\ref{eq:tGAS}) as a data generating process for log-returns. After computing log-prices, we randomly censor them to mimic asynchronous trading. We then estimate both the local-level model with score-driven covariances and the $t$-GAS. There are two possible methods to estimate the $t$-GAS in presence of missing values. The first consists in synchronizing data through an interpolation scheme, e.g. the previous-tick. We have seen in the previous application that this generally leads to a downward bias. The second method consists in computing the score with respect to the marginal density of observed data (\citealt{LUCAS201696}). In the case of the multivariate Student-$t$ distribution, this is particularly simple, since it is known that the marginals are Student-$t$ densities with the same number of degrees of freedom. Specifically, assume $X\sim t_{\nu}(0,\Sigma)\in\mathbb{R}^n$ and partition the elements in $X$, $\Sigma$ as:
 \begin{equation}
  X =
  \begin{pmatrix}
   X_1\\
   X_2
  \end{pmatrix},
  \quad
  \Sigma = 
  \begin{pmatrix}
   \Sigma_{11}, \Sigma_{12}\\
   \Sigma_{21}, \Sigma_{22}
  \end{pmatrix}
 \end{equation}
where $X_1\in\mathbb{R}^p$, $p<n$ and $\Sigma_{11}\in\mathbb{R}^{p\times p}$. The marginal density of $X_1$ is then a $t_{\nu}(0,\Sigma_{11})$. See e.g. \cite{kibria} for details. Missing values can therefore be handled as in the Kalman filter, i.e. by computing the score of the marginal density $t_{\nu}(0,\Gamma_t\Sigma_t\Gamma_t')$, where $\Gamma_t$ is a selection matrix with ones in the columns corresponding to traded assets.

This method avoids the introduction of artificial zero returns. However, it induces significant data reduction, as only consecutive trades can be used to compute returns. For instance, if the $i$-th assets is traded at time $t$, but is not traded at time $t-1$, the return $r_{i,t}$ is treated as a missing value and the information related to the price at time $t$ is lost. The local-level instead exploits the observation of the price at time $t$ to reconstruct the vector of efficient prices and to update covariances. 

To illustrate this concept, we simulate log-returns through a $t$-GAS model with $n=10$ and, in order to emphasize the effect of fat-tails, we set $\nu=3$. Ten different scenarios are considered, characterized by an increasing level of asynchronicity ($\lambda=0,0.1,\dots,0.9$). We perform $n=250$ simulations of time-series of $T=4000$ observations. As in the previous study, in-sample estimates are computed in the first subsample of $T_{\text{sub}}=2000$ observations, while out-of-sample estimates are computed in the second subsample. As a loss measure, we use the Frobenius distance, defined as:
\begin{equation}
 ||\hat{\Sigma}_t-\Sigma_t||_\text{F}=\sqrt{\text{Tr}[(\hat{\Sigma}_t-\Sigma_t)^2]}
\end{equation}
The parameterization of $R_t$ used in both models is the one based on hyperspherical coordinates, and the static parameters driving the dynamic elements in $D_t$ and $R_t$ have the restrictions in eq. (\ref{eq:ThetaUnrestr}). 
%
%
%
\begin{figure}
\centering
\begin{minipage}{0.5\textwidth}
\centering
    \includegraphics[width=1\linewidth]{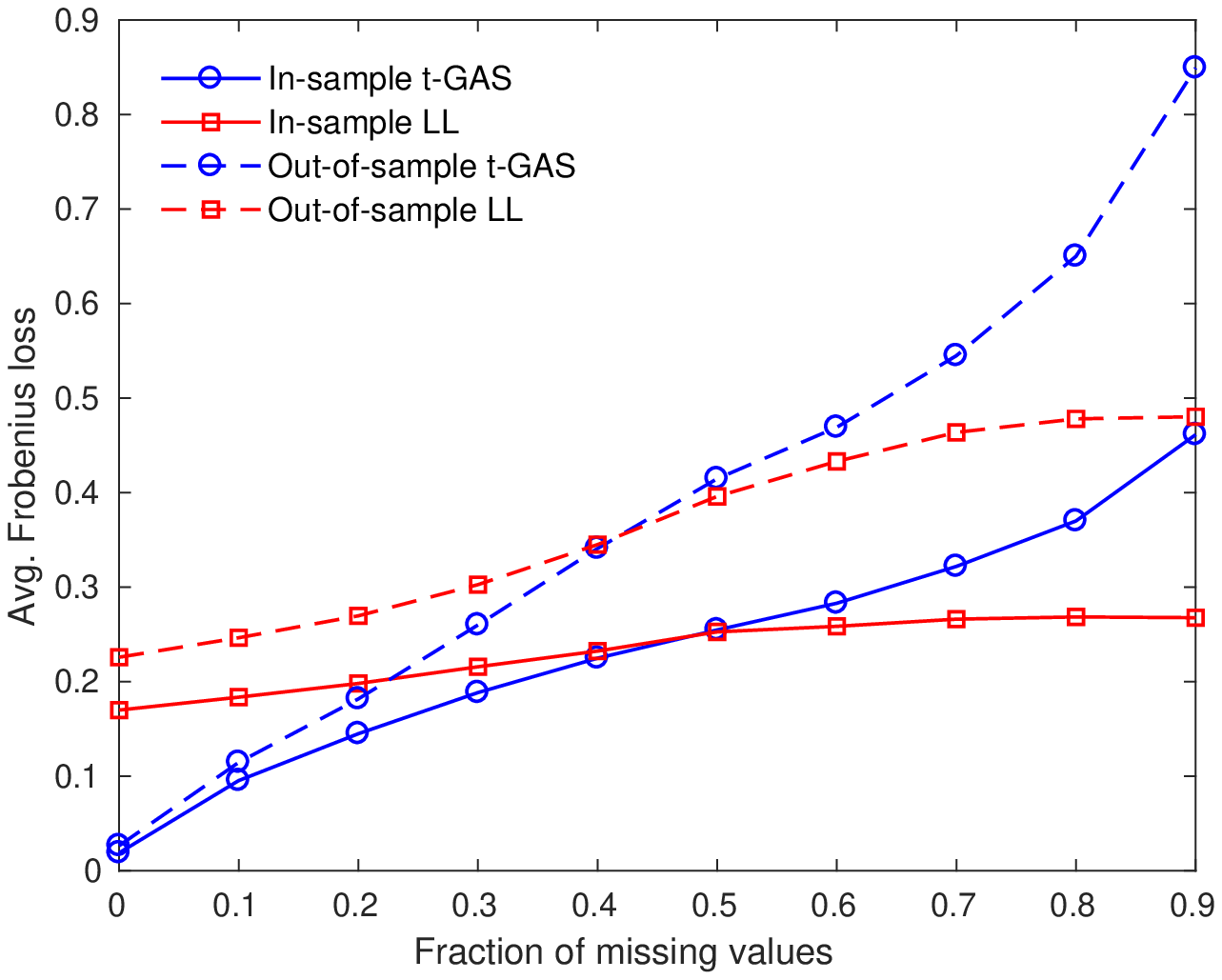}
\end{minipage}%
\begin{minipage}{0.5\textwidth}
\centering
    \includegraphics[width=1\linewidth]{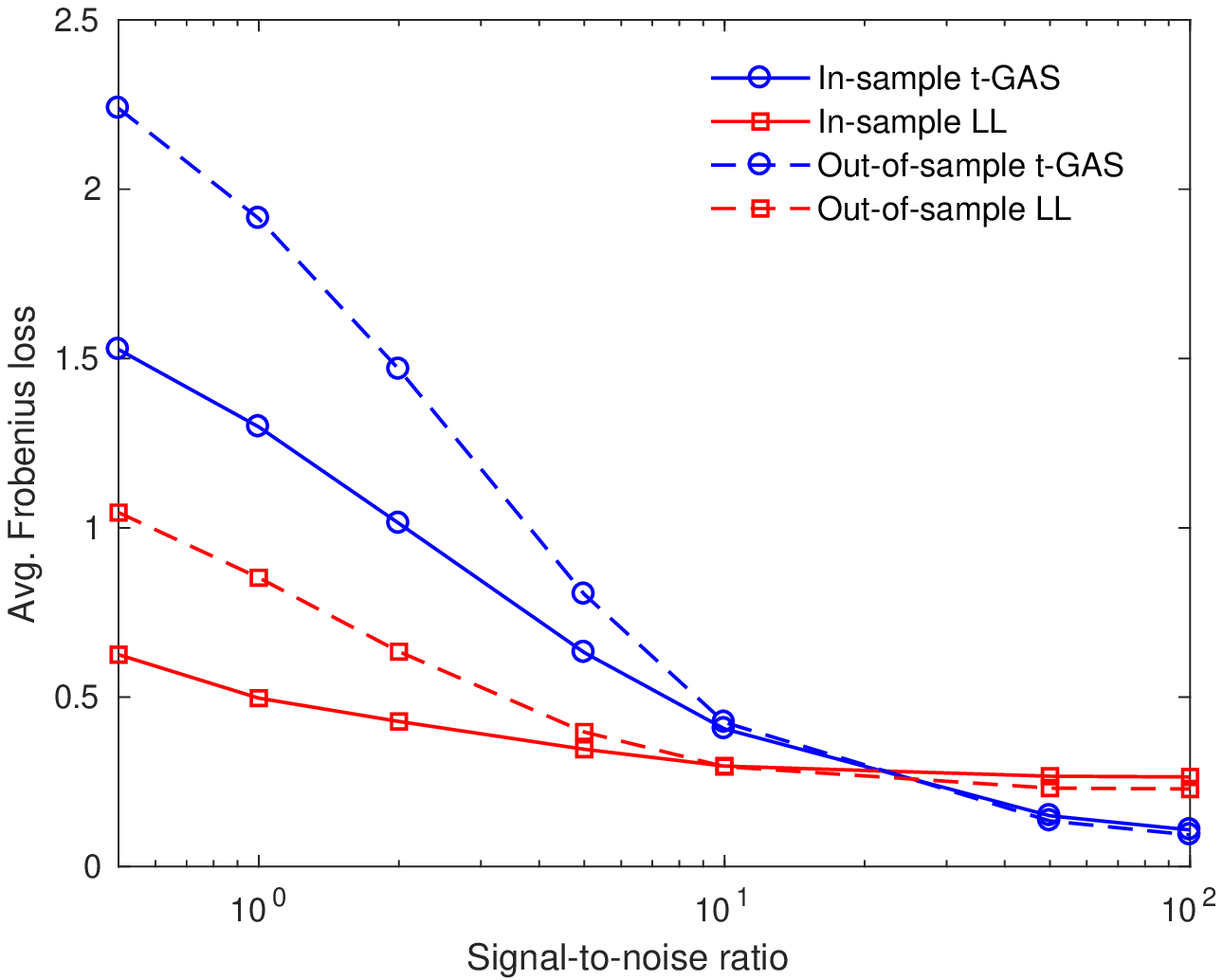}
\end{minipage}
    \caption{Left: in-sample and out-of-sample Frobenius losses of $t$-GAS and local-level model as a function of the probability of missing values $\lambda$. The DGP is a $t$-GAS with $\nu=3$. Right: in-sample and out-of-sample Frobenius losses of $t$-GAS and local-level model as a function of the signal-to-noise ratio. The DGP is a $t$-GAS with $\nu=3$ contaminated by a Student-$t$ distributed measurement error with $\nu_{err}=3$. }
    \label{fig:simIncAsynTgas}
\end{figure}

Figure \ref{fig:simIncAsynTgas} shows on the left the results of the simulation study. There is an evident trade-off between fat-tails and asynchronicity. Not surprisingly, if data are perfectly synchronized, the $t$-GAS model performs better. However, once missing values come into play, the local-level model tends to improve its relative performance. In particular, for $\lambda > 0.5$, it provides better in-sample and out-of-sample estimates. Note that the average losses of $t$-GAS rapidly increase with $\lambda$. For $\lambda=0.9$, the in-sample loss of $t$-GAS is $\sim 24$ times larger than the one at $\lambda=0$, while for the local-level it is just $\sim 1.7$ times larger. The rapid deterioration of the relative performance of the $t$-GAS in presence of missing values is due to the aforementioned effect of data reduction. A similar tradeoff is observed when we add a noise to the simulated prices. In a second experiment, we ignore asynchronicity, but contaminate the prices simulated through the $t$-GAS with a zero-mean Student-$t$ distributed measurement error with $\nu_{err}=3$. We consider different levels of signal-to-noise ratio, from $\delta=0.1$ to $\delta=100$. As shown on the right in figure \ref{fig:simIncAsynTgas}, for $\delta<20$, the average loss of the local-level model is significantly lower that that of the $t$-GAS, both in-sample and out-of-sample. This is due to the fact that, in the local-level, we model the covariances of latent returns, rather than those of observed returns. Of course, adding noise and asynchronicity together would lead to further increase the relative difference between the two models.  

The results of this analysis suggest that the choice of the model depends on the tradeoff between outliers and asynchronicity/noise. As will be evident in our empirical application, high-frequency data are characterized by high levels of asynchronicity, with $\lambda>0.8$, and are significantly noisy, having $\delta<1$. The use of the local-level model is therefore more suited in an high-frequency setting, where asynchronicity and microstructure effects are extremely important.  

\section{Empirical illustration}
\label{sec:empIll}

\subsection{Dataset}
Our dataset contains 1-second transaction data of 10 most frequently traded NYSE assets in 2014. Data are related to transactions from 02-01-2014 to 31-12-2014, including a total of 251 business days. The exchange opens at 9.30 and closes at 16.00 local time, so that the number of seconds per day is $T=23400$. We perform the standard procedures described by \cite{BN2} to clean the data. Table \ref{tab:dataset} shows summary statistics of the 10 assets, including the average duration (in seconds) between trades, the average probability of missing values $\lambda_{\text{avg}}$ (computed as the average fraction of trades per day) and the average number of trades per day. We also report the average signal-to-noise ratio $\delta_{\text{avg}}$, computed from the estimated matrices $D_t$, $H_t$. Specifically, $\delta_{\text{avg}}$ is defined as the ratio $D_{t,ii}^2/H_{t,ii}$, $i=1,\dots,10$, averaged over all timestamps. 

The numbers in the table provide a first evidence about the relevance of asynchronous trading and microstructure effects in high-frequency data. The average probability of missing values is greater than 0.8, indicating large levels of sparsity even for the liquid assets included in this dataset. The average signal-to-noise is lower than 1 for most of the assets. This implies that also microstructure effects play a relevant role. As done in the simulation study in Section \ref{sec:MC}, we thus aim to show the advantages of the proposed method over standard techniques ignoring these effects.

\bigskip
\begin{table}[!htbp]
\centering
\small
\setlength{\tabcolsep}{12pt}
 \begin{tabular}{c|ccccc} 
 \hline
\multicolumn{1}{c|}{Symbol} & Asset & Avg. duration (sec) & $\lambda_{\text{avg}}$ & Avg. n. of trades & $\delta_{\text{avg}}$ \\ 
\hline
XOM              & Exxon Mobil Corporation   &  5.434 & 0.816 & 4304  & 1.178 \\                                
C                & Citigroup Inc   &  6.135 & 0.836 & 3832  & 1.246   \\
JPM              & JPMorgan Chase   &  6.250 & 0.840 & 3743  & 0.999  \\ 
HAL              & Halliburton Company   &  6.369 & 0.843 & 3690  & 0.872  \\
CVX              & Chevron   &  6.579 & 0.848 & 3553  & 0.850   \\
DIS              & Walt Disney   &  6.622 & 0.849 & 3543  & 0.846   \\  
JNJ              & Johnson \& Johnson   &  6.666 & 0.850 & 3529  & 0.809  \\
SLB              & Schlumberger Limited   &  6.802 & 0.853 & 3454  & 0.613   \\
DAL              & Delta Air Lines   &  6.993 & 0.857 & 3348  & 0.766   \\
WMT              & Walmart   &  7.042 & 0.858 & 3325  & 0.698  \\
\hline
\end{tabular}
\caption{Summary statistics and average signal-to-noise ratio of high-frequency transaction data employed in the empirical analysis.}
\label{tab:dataset}
\end{table}

\subsection{Structure of static parameters}
\label{sub:structStatic}

\begin{table}[!htbp]
\centering
\small
\setlength{\tabcolsep}{12pt}
 \begin{tabular}{ccccccc} 
 \hline
Restriction type & 23 Jan & 11 Apr & 27 May & 6 Aug & 4 Sep & 16 Oct\\ 
\hline
$I$    &  $-3.5268$        & $-3.7727$  & $-2.3330$ & $-2.3193$ & $-2.4435$ & $-5.5497$ \\      
$II$   & $-3.6014$         & $-3.7914$  & $-2.3787$ & $-2.3765$ & $-2.5160$ & $-5.5502$ \\
\hline
\end{tabular}
\caption{AIC ($\times 10^5$) statistics computed on five randomly selected trading days with the parameter restriction $I$ based on eq. (\ref{eq:ThetaUnrestr}) and the restriction $II$ based on eq. (\ref{eq:ThetaRW}). }
\label{tab:paramAic}
\end{table}

The vector $\omega$ and the matrices $A$, $B$ in the dynamic equation (\ref{eq:gas}) have dimensions $k\times 1$ and $k\times k$, respectively. For $n=10$, $k$ is equal to $21$ in the equicorrelation parameterization, while it is equal to $65$ in the parameterization based on hyperspherical coordinates. It is therefore necessary to impose restrictions on the parameter space in order to estimate the model. We perform an AIC test to choose among different restrictions. Specifically, we compare the parameter structure $I$ in eq. (\ref{eq:ThetaUnrestr}) to the parameter structure $II$ in eq (\ref{eq:ThetaRW}) Table \ref{tab:paramAic} reports AIC statistics of the local-level model estimated on five randomly selected trading days with the two restrictions. The parameterization used in this test is the one based on hyperspherical coordinates, however similar results are obtained with the equicorrelation parameterization. The AIC provided by the random walk-type restriction $II$ is always lower. As will be shown in next sections, the intuitive reason behind this result is that intraday covariances have non-stationary dynamics and do not mean revert to an unconditional level. We therefore implement this restriction when estimating the local-level model in all subsequent analyses.  

\subsection{In-sample analysis}

%

\begin{figure}
\centering
  \begin{minipage}{0.5\textwidth}
    \includegraphics[width=1\linewidth]{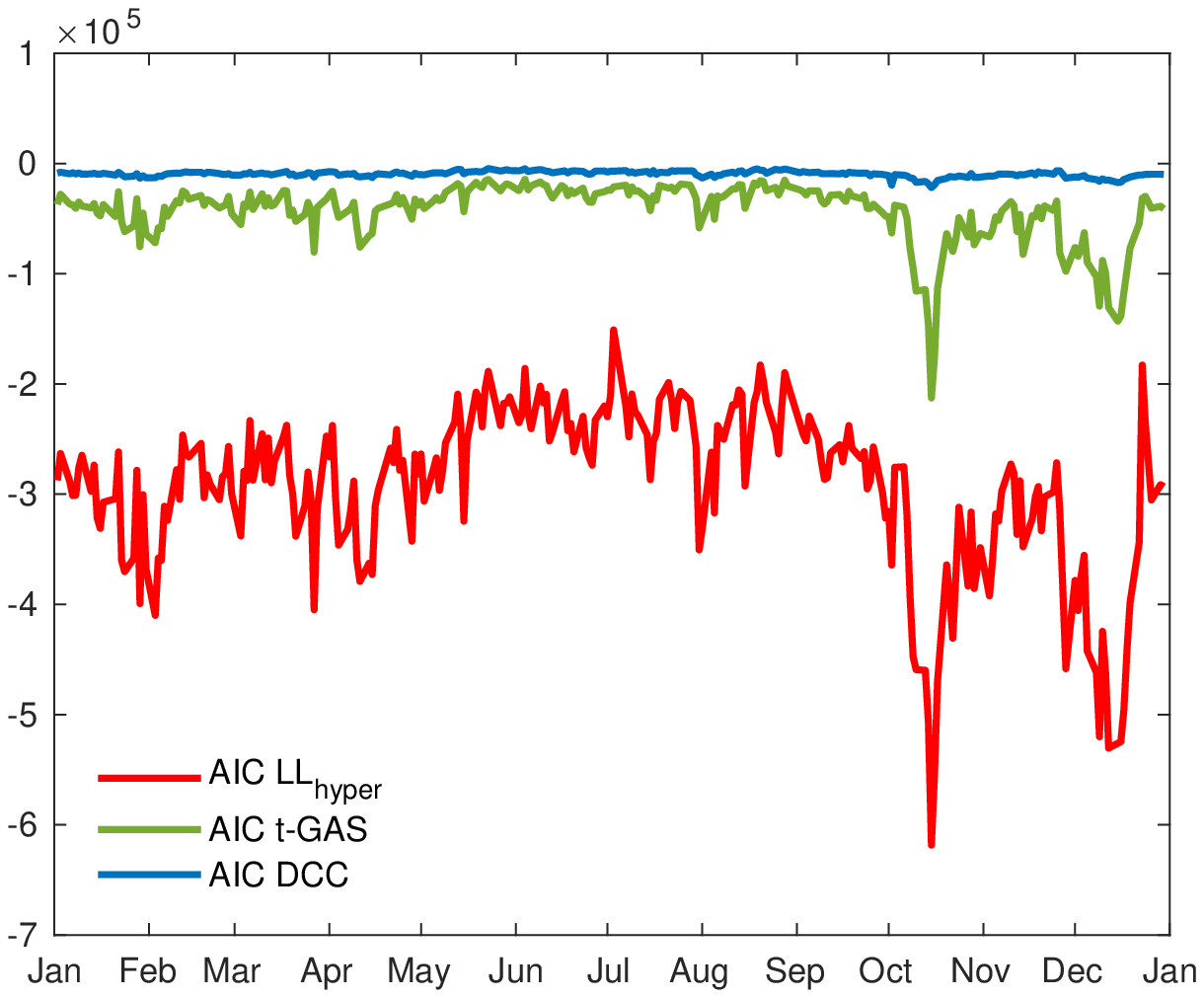}
    \end{minipage}%
\begin{minipage}{0.5\textwidth}
\centering
    \includegraphics[width=1\linewidth]{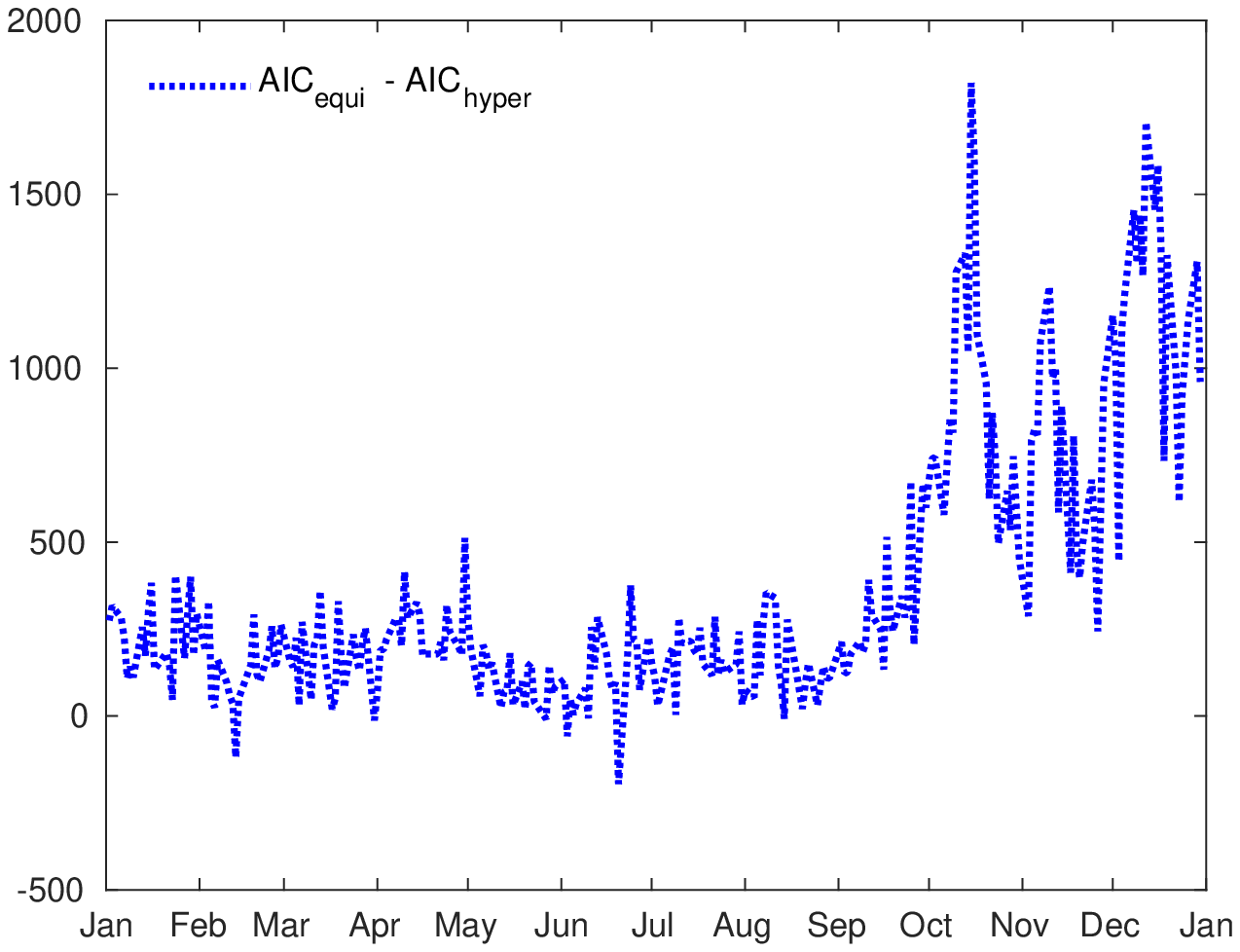}
    \end{minipage} 
    \caption{Left: AIC of local-level with hyperspherical coordinates, $t$-GAS and DCC. Right: differences between AIC of local-level with equicorrelation parameterization and local-level with hyperspherical coordinates.}
    \label{fig:aicAnalysis}
\end{figure}

We estimate the local-level model on each business day of 2014 (the average estimation time on Matlab 2018 is 28.3 min on an i7-2600 CPU at 3.40GHz). In order to initialize time-varying parameters, we proceed as in Section 3.1, i.e. we estimate the static version of the model in a pre-sample including the first 15 minutes of the trading day. The starting values of time-varying parameters are then set equal to the static estimates. As a parameterization for correlations, we use both hyperspherical coordinates and equicorrelations. We compare covariance estimates obtained through the local-level model to the covariances of the $t$-GAS model. As in \cite{GAS2}, correlations in the $t$-GAS are parameterized using hyperspherical coordinates. In presence of asynchronous data, the $t$-GAS can be estimated by previous-tick synchronization or using the missing value approach of \cite{LUCAS201696}. As discussed in Sections \ref{sub:MC:noiseAyn}, \ref{sub:MC:outliers}, the first method leads to a downward bias, while the second implies data reduction, as a consequence of modelling returns rather than prices. We use the missing value approach, as zero returns dramatically jeopardize the estimation of correlations. The same random walk-type restrictions on static parameters are implemented for the $t$-GAS. As an additional benchmark, we use the DCC model estimated through previous-tick interpolation. When data are sampled at the largest available frequency, the correlations provided by the DCC turn out to be flat and close to zero. This is due to both microstructure effects and zero returns. In order to attenuate these two effects, we aggregate data at a lower frequency. We use a sampling frequency of 20 seconds. Indeed, at higher frequencies correlations are significantly biased, while at lower frequencies the inference becomes extremely inefficient due to data reduction. 

\begin{figure} 
\centering
    \begin{minipage}{0.5\textwidth}
    \centering
    \includegraphics[width=1\linewidth]{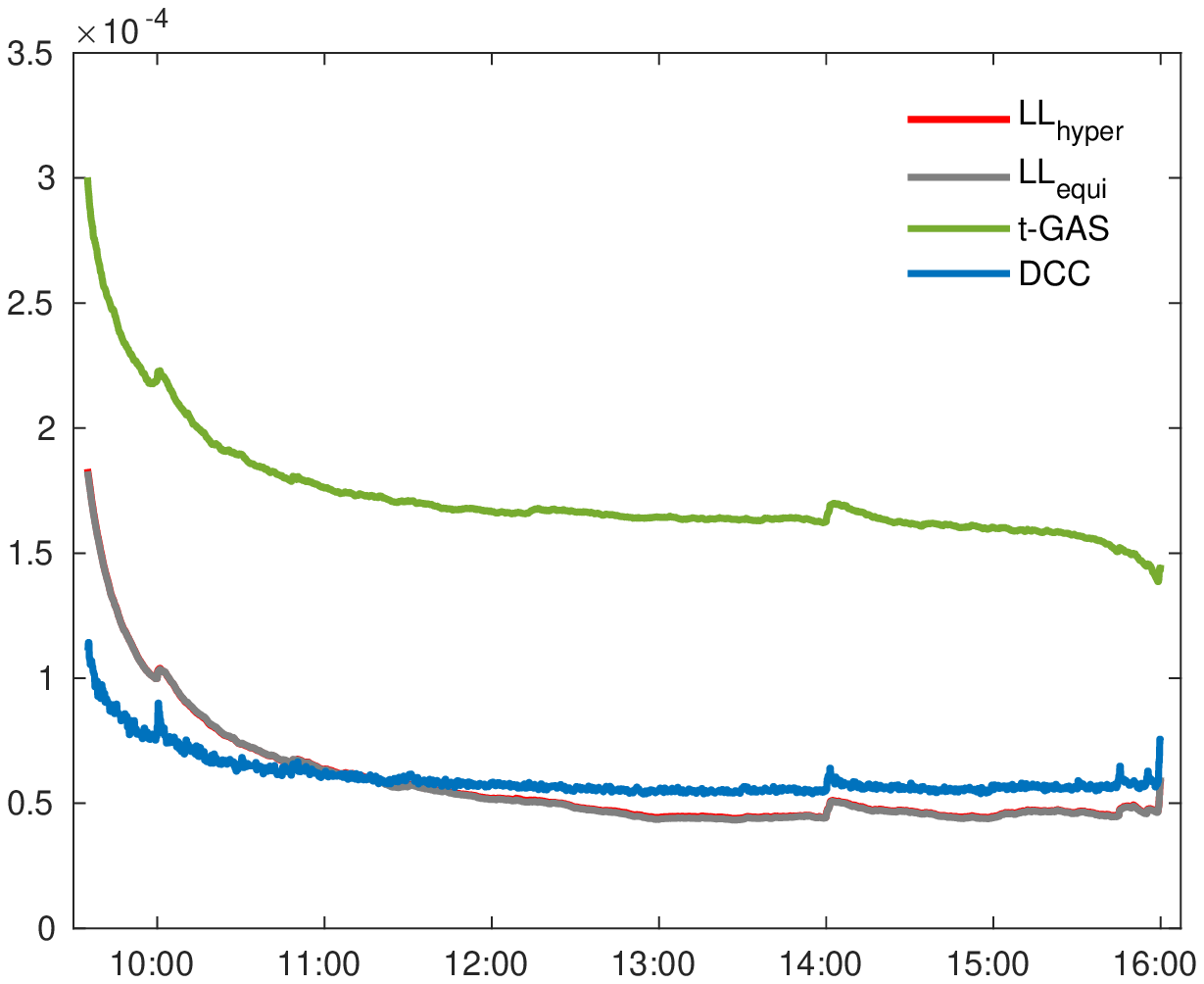}
    \end{minipage}%
\begin{minipage}{0.5\textwidth}
\centering
    \includegraphics[width=1\linewidth]{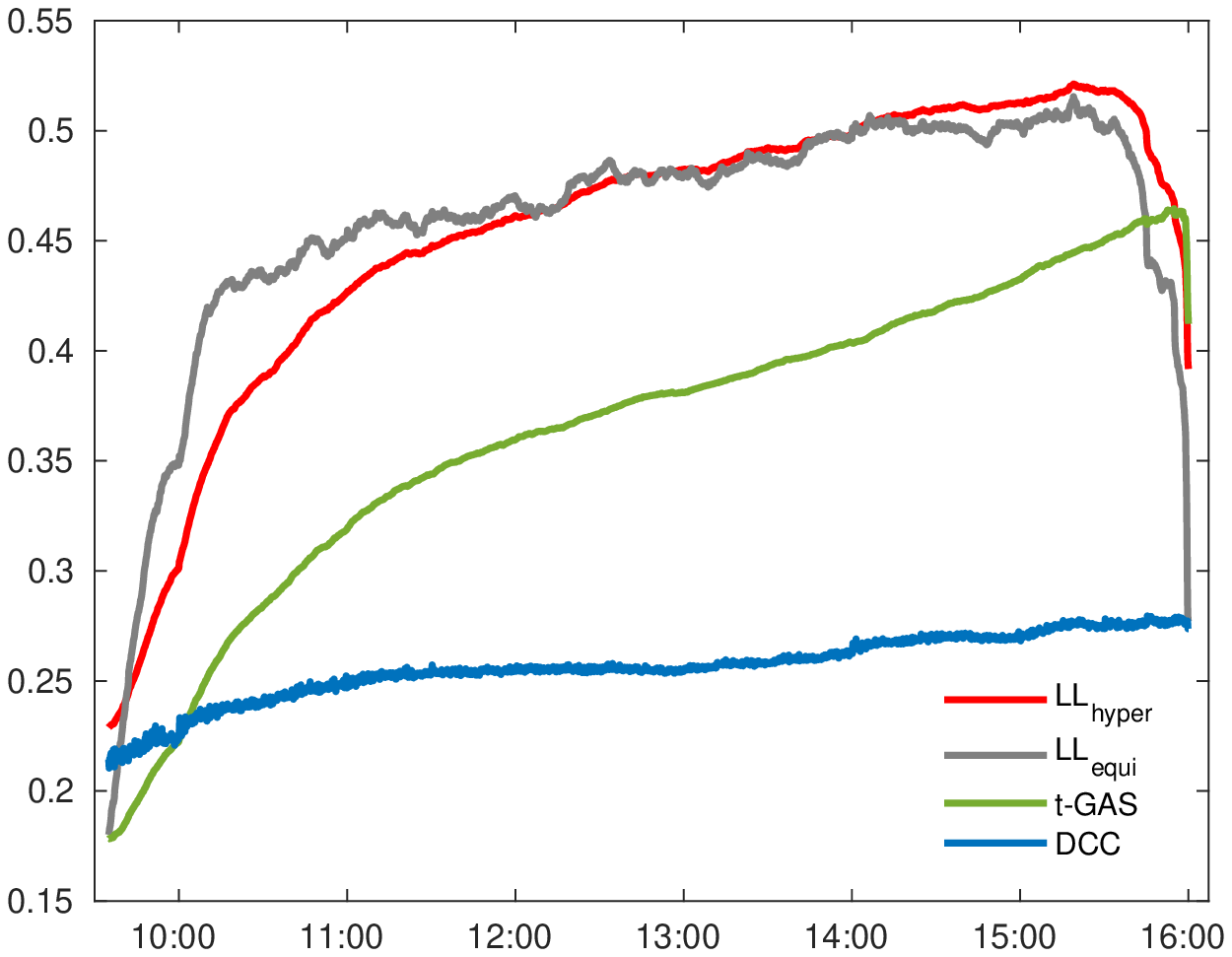}
    \end{minipage} 
    \caption{Intraday average volatilities (left) and correlations (right) of local-level with hyperspherical coordinates, local-level with equicorrelation parameterization, $t$-GAS and DCC. Volatilities are average over all the 10 assets while correlations are averaged over all couples of assets.}
    \label{fig:intraDayCorr}
\end{figure}

Figure \ref{fig:aicAnalysis} shows on the left the daily AIC provided by the local-level with hyperspherical coordinates, the $t$-GAS and the DCC. The AIC of the local-level model is significantly lower than that of $t$-GAS and DCC. These differences are mainly due to data reduction and provide a quantitative assessment of the relevance of this effect. On the right, we plot the daily difference between the AIC of the equicorrelation parameterization and the AIC of hyperspherical coordinates. The difference is positive in most of the days of the sample, indicating better in-sample fit for hyperspherical coordinates. However, note that the relative difference is small, meaning that the loss due to the equicorrelation assumption does not affect significantly the quality of the fit. 

Figure \ref{fig:intraDayCorr} shows the estimated intraday volatilities and correlations averaged over the whole sample of 251 business days. Volatilities are averaged over the 10 assets, while correlations are averaged over all couples of assets. The intraday pattern of volatilities provided by the $t$-GAS is higher than that of the two local-level models. The reason is that the estimated volatilities in the $t$-GAS are affected by microstructure noise, while the local-level sets the efficient return volatilities in $D_t$ apart from the noise variances in $H_t$. Microstructure effects thus lead to larger variances in the $t$-GAS. The average volatility estimated by the DCC is closer to that of the two-local level models. Indeed, microstructure effects in the DCC are attenuated by sampling at a lower frequency. However, previous-tick interpolation leads to correlations which are significantly biased toward zero. The average correlations of the $t$-GAS are significantly less biased compared to DCC correlations, confirming the ability of the missing value approach to avoid the distortions due to the introduction of artificial zero returns. Nevertheless, they suffer from the additional bias due to microstructure effects. Overall, the in-sample analysis provides clear evidence that the proposed approach is more suited when dealing with intraday covariances. 

The intraday patterns of the local-level model with hyperspherical coordinates are close to those provided by the equicorrelation parameterization, in accordance with the result in figure \ref{fig:aicAnalysis}. Larger deviations are observed in the correlations, in particular in the first part of the trading day, where the average equicorrelation increases at a slightly higher rate, and during the last few minutes, where the decrease is more pronounced. 

\subsection{Intraday dynamics of covariances}

\begin{figure}[ht] 
\centering
  \begin{minipage}{0.5\textwidth}
  \centering
    \includegraphics[width=1\linewidth]{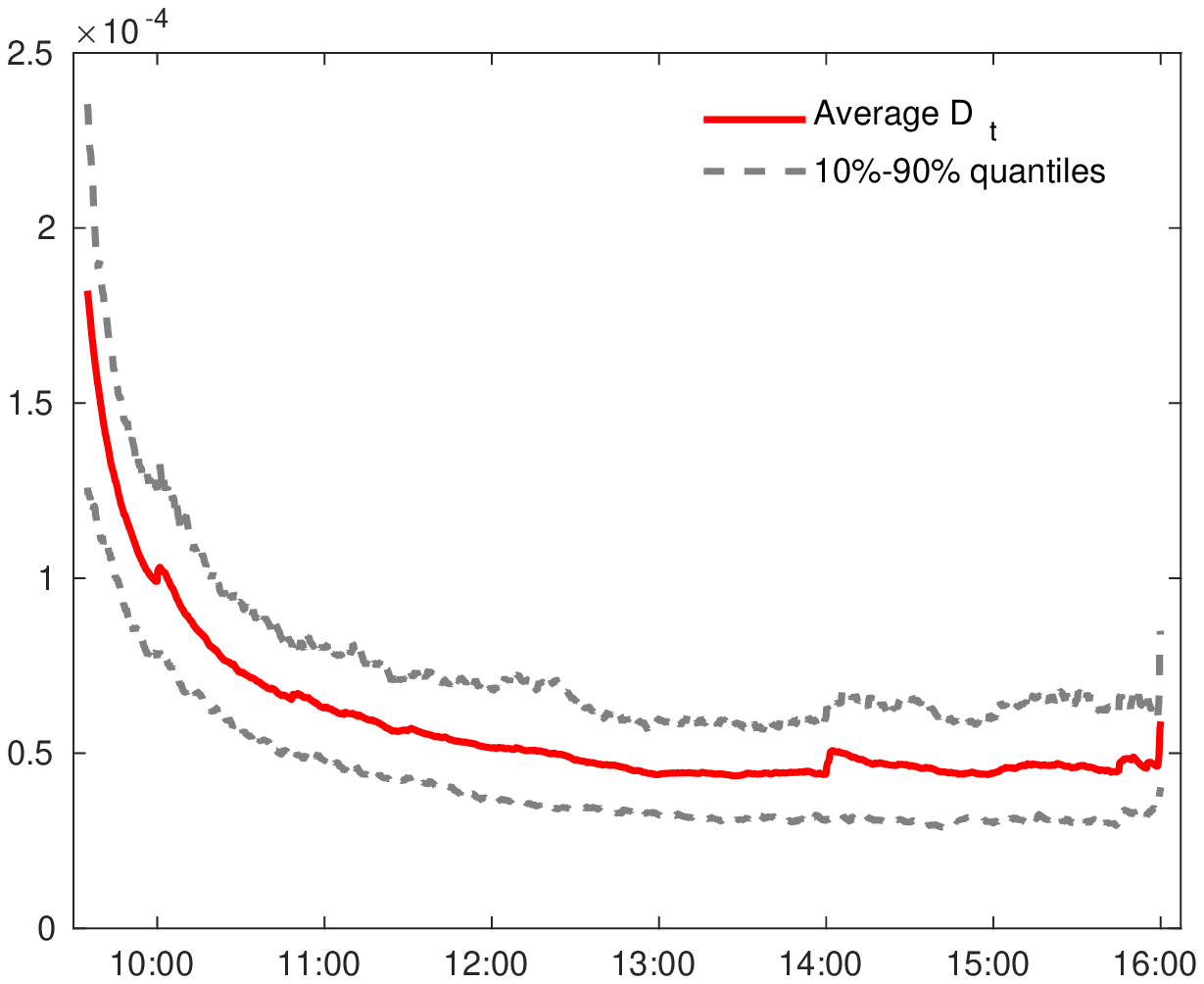} 
  \end{minipage}%
  \begin{minipage}{0.5\textwidth}
  \centering
    \includegraphics[width=1\linewidth]{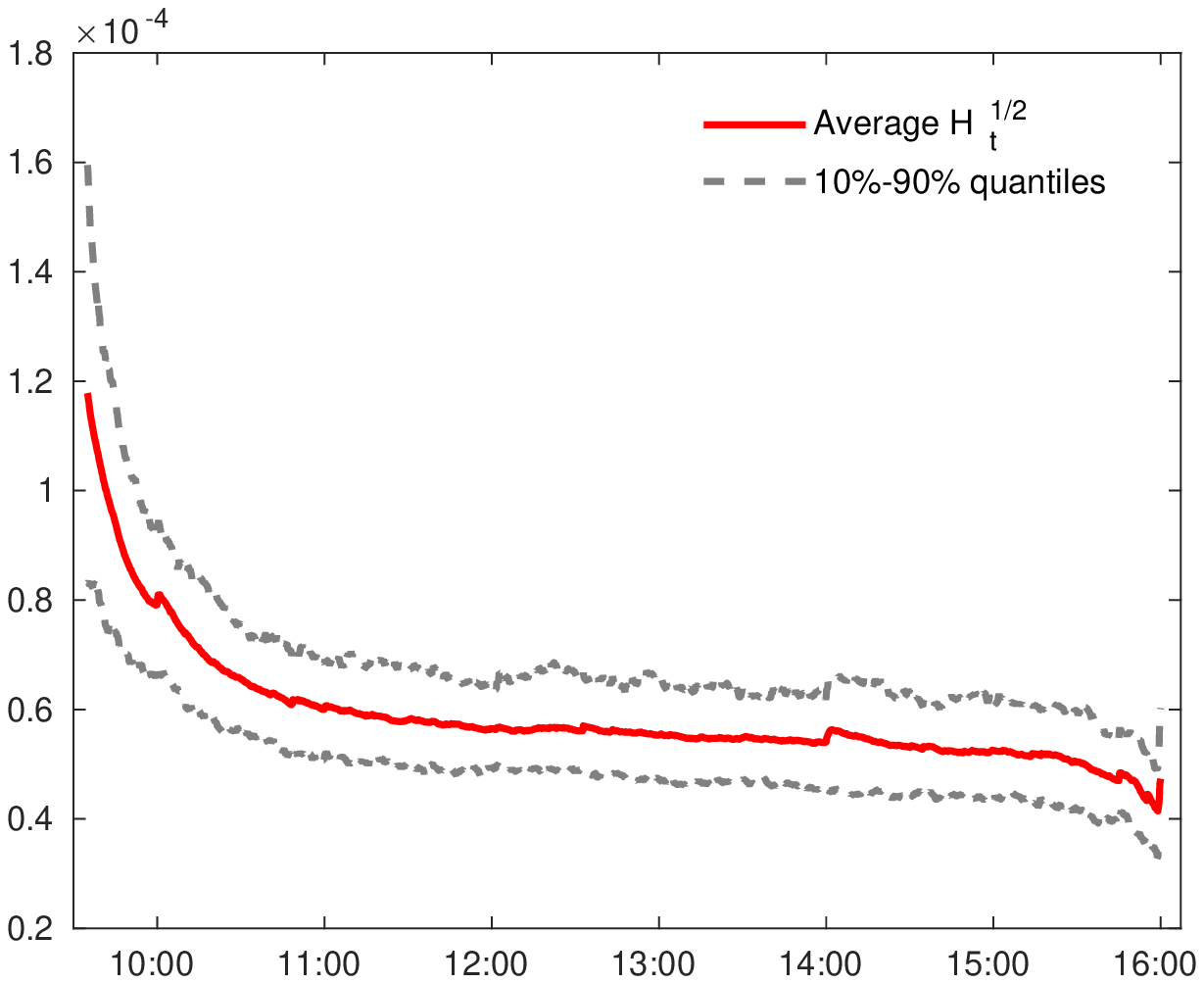} 
  \end{minipage} 
  \begin{minipage}{0.5\textwidth}
  \centering
    \includegraphics[width=1\linewidth]{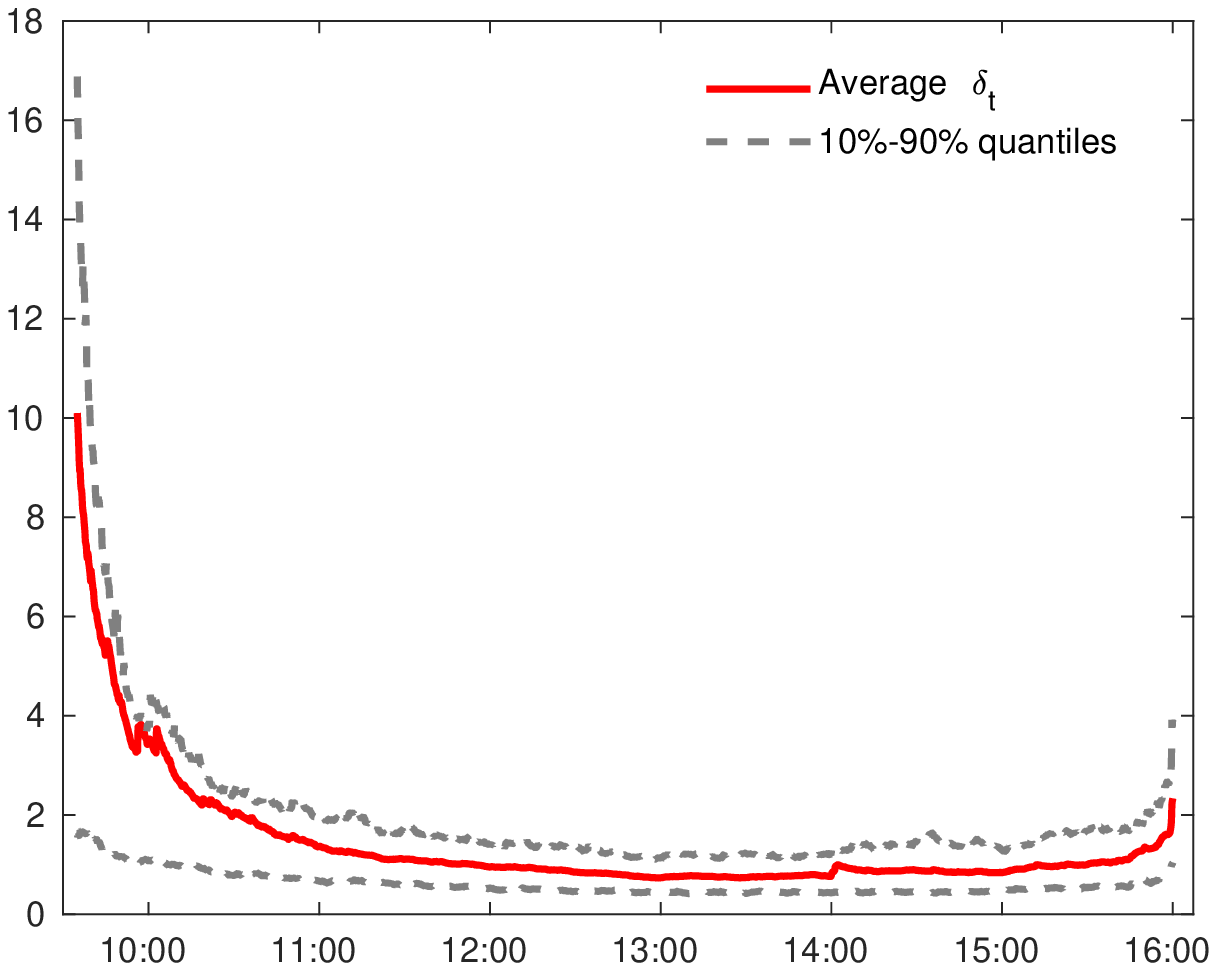} 
  \end{minipage}%
  \begin{minipage}{0.5\textwidth}
  \centering
    \includegraphics[width=1\linewidth]{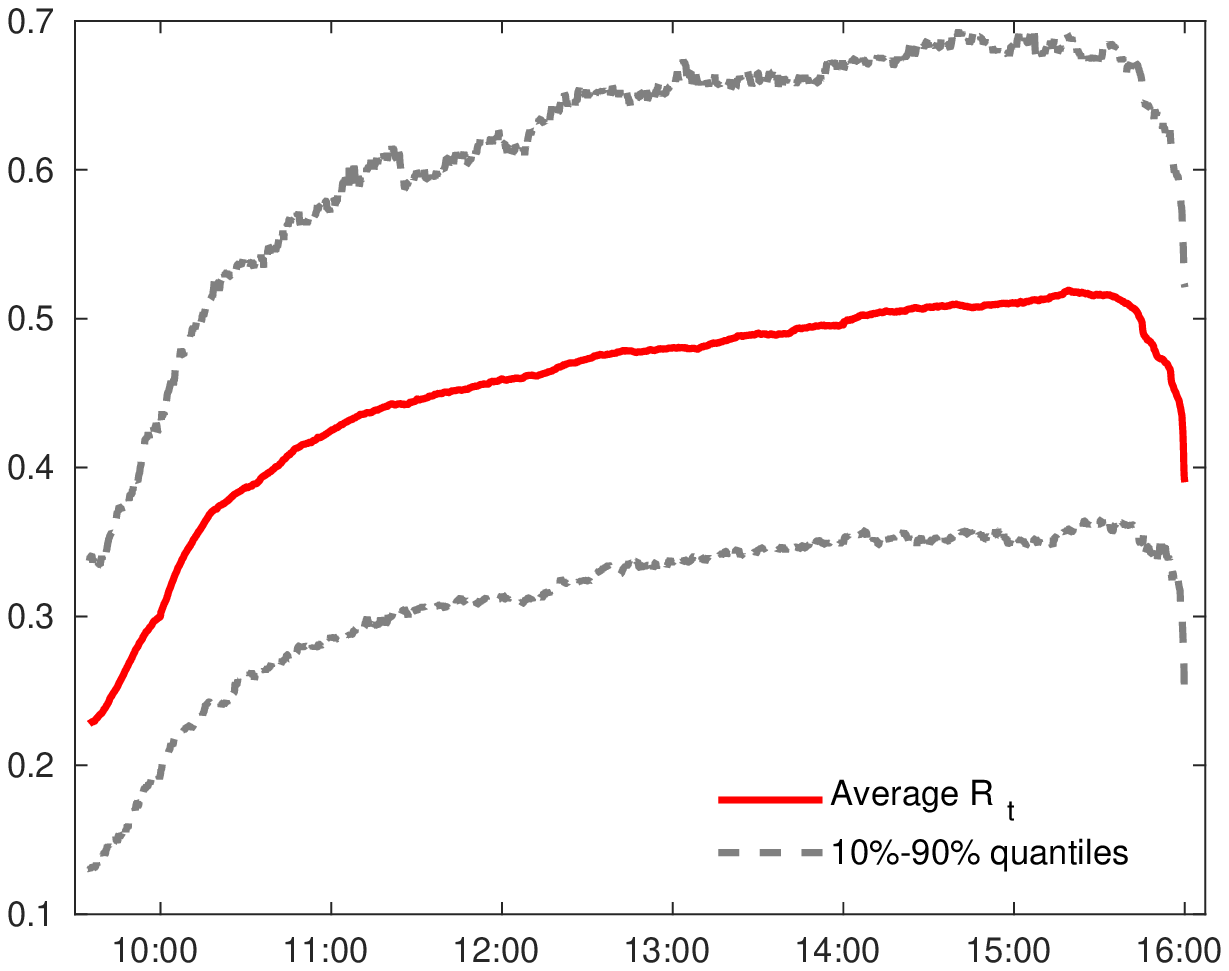} 
  \end{minipage} 
  \caption{Average intraday patterns of the estimated time-varying parameters $D_t$, $H_t^{1/2}$, $\delta_t$ and $R_t$. For each $t=1,\dots,23400$, we report 10\% and 90\% quantiles of the distribution (over days) of averages of $D_t$, $H_t^{1/2}$, $\delta_t$ and $R_t$ (over assets or couples of assets).}
    \label{fig:tvTime} 
\end{figure}

\begin{figure}[ht] 
\centering
  \begin{minipage}{0.5\textwidth}
  \centering
    \includegraphics[width=1\linewidth]{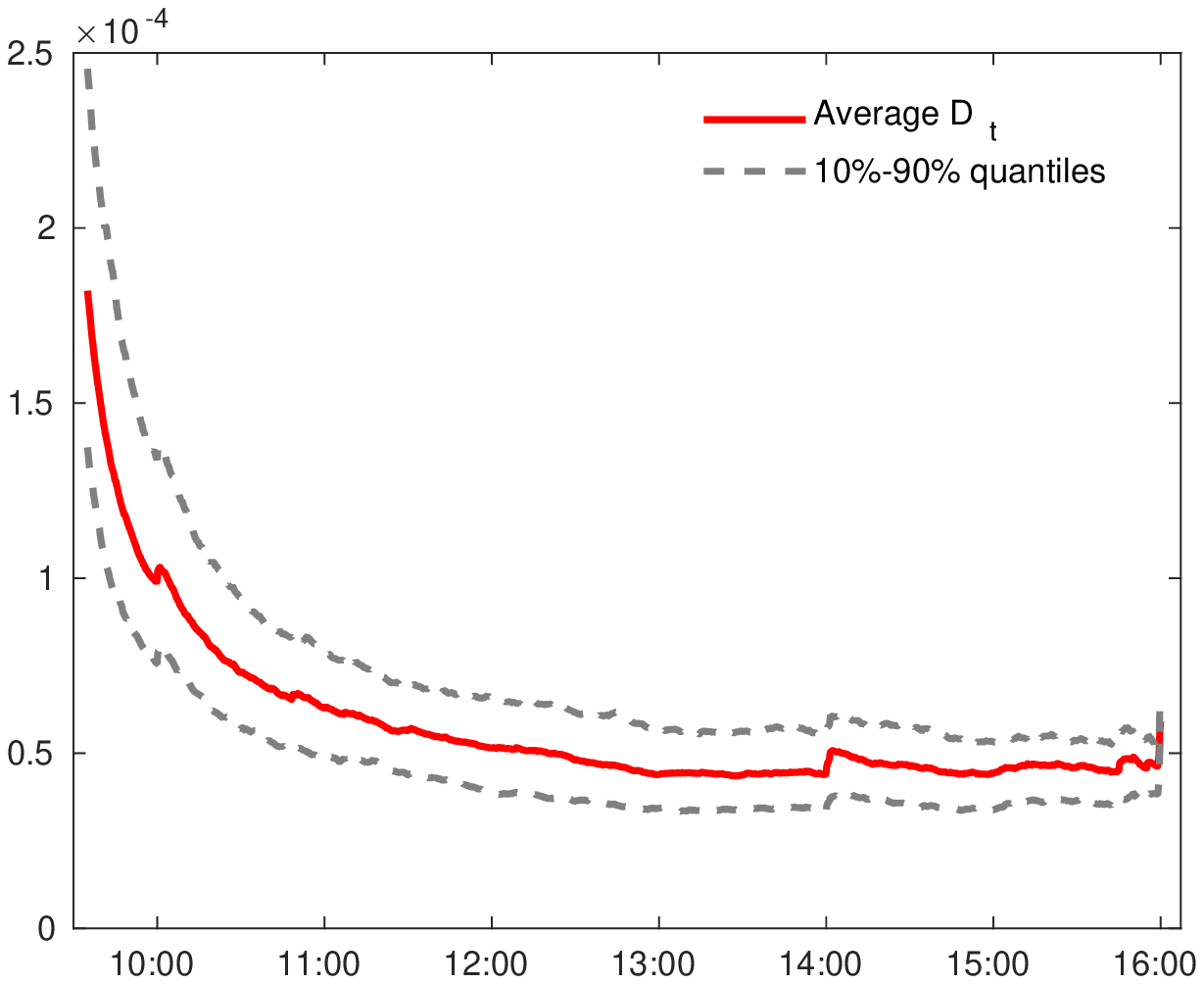} 
  \end{minipage}%
  \begin{minipage}{0.5\textwidth}
  \centering
    \includegraphics[width=1\linewidth]{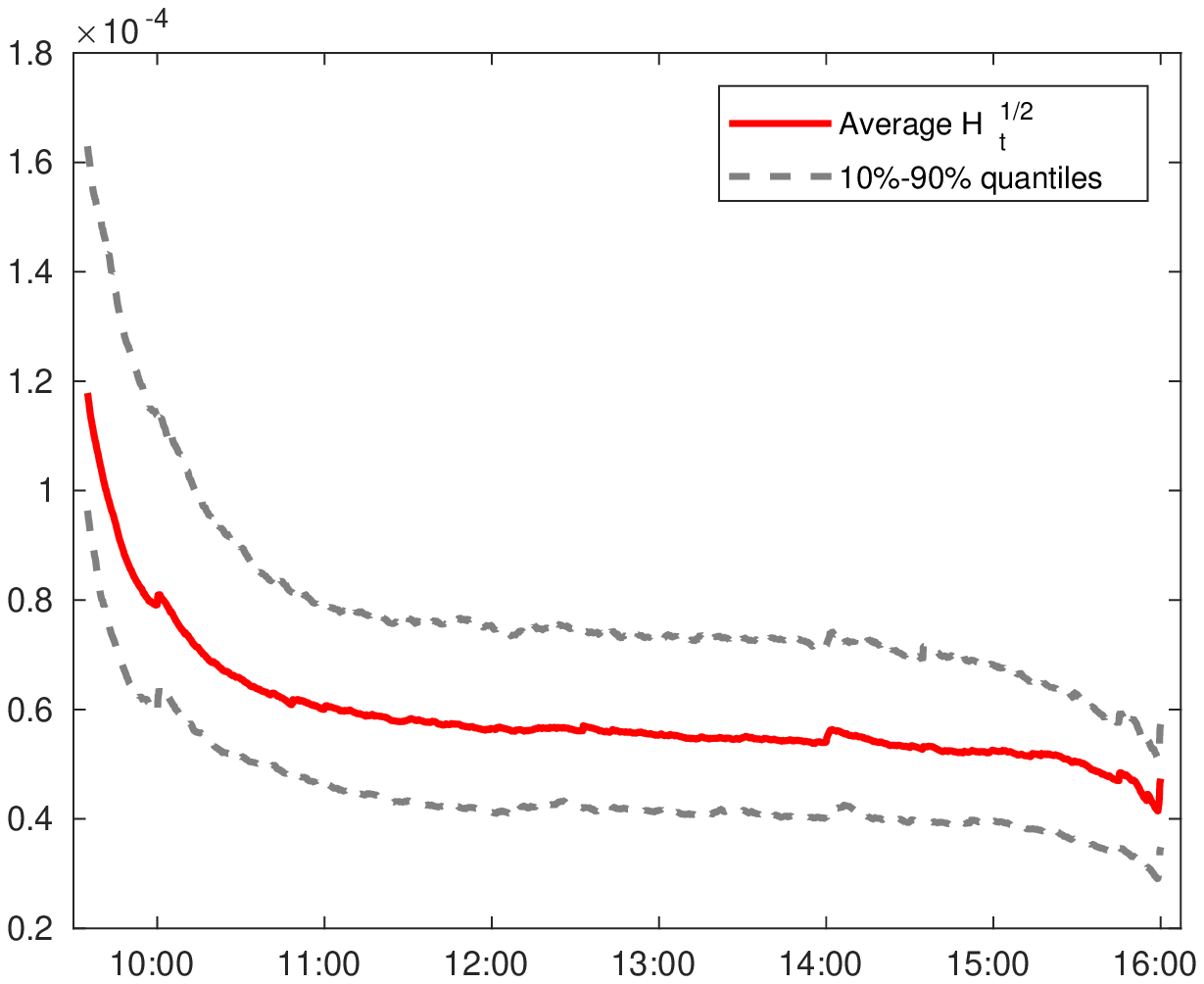} 
  \end{minipage} 
  \begin{minipage}{0.5\textwidth}
  \centering
    \includegraphics[width=1\linewidth]{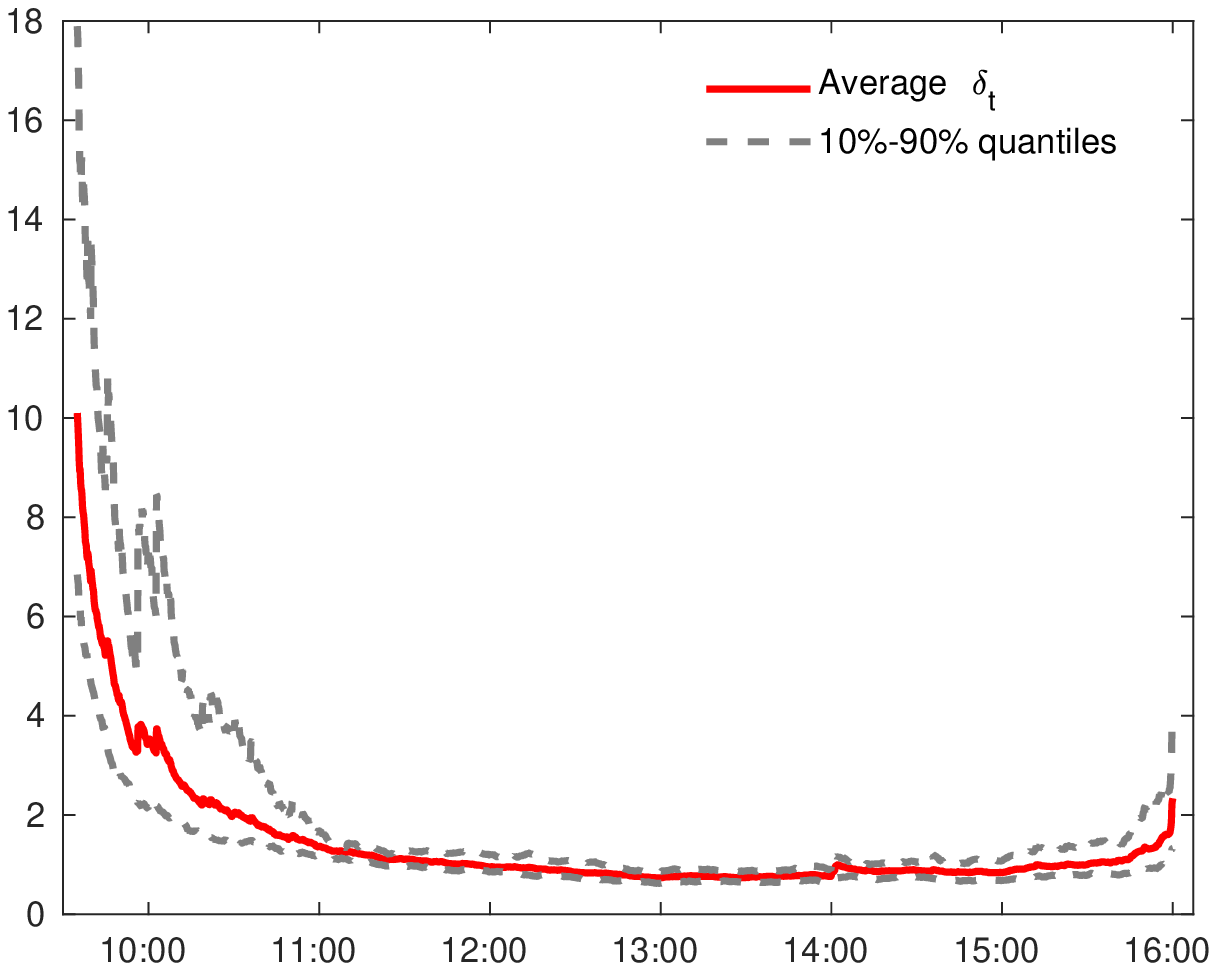} 
  \end{minipage}%
  \begin{minipage}{0.5\textwidth}
  \centering
    \includegraphics[width=1\linewidth]{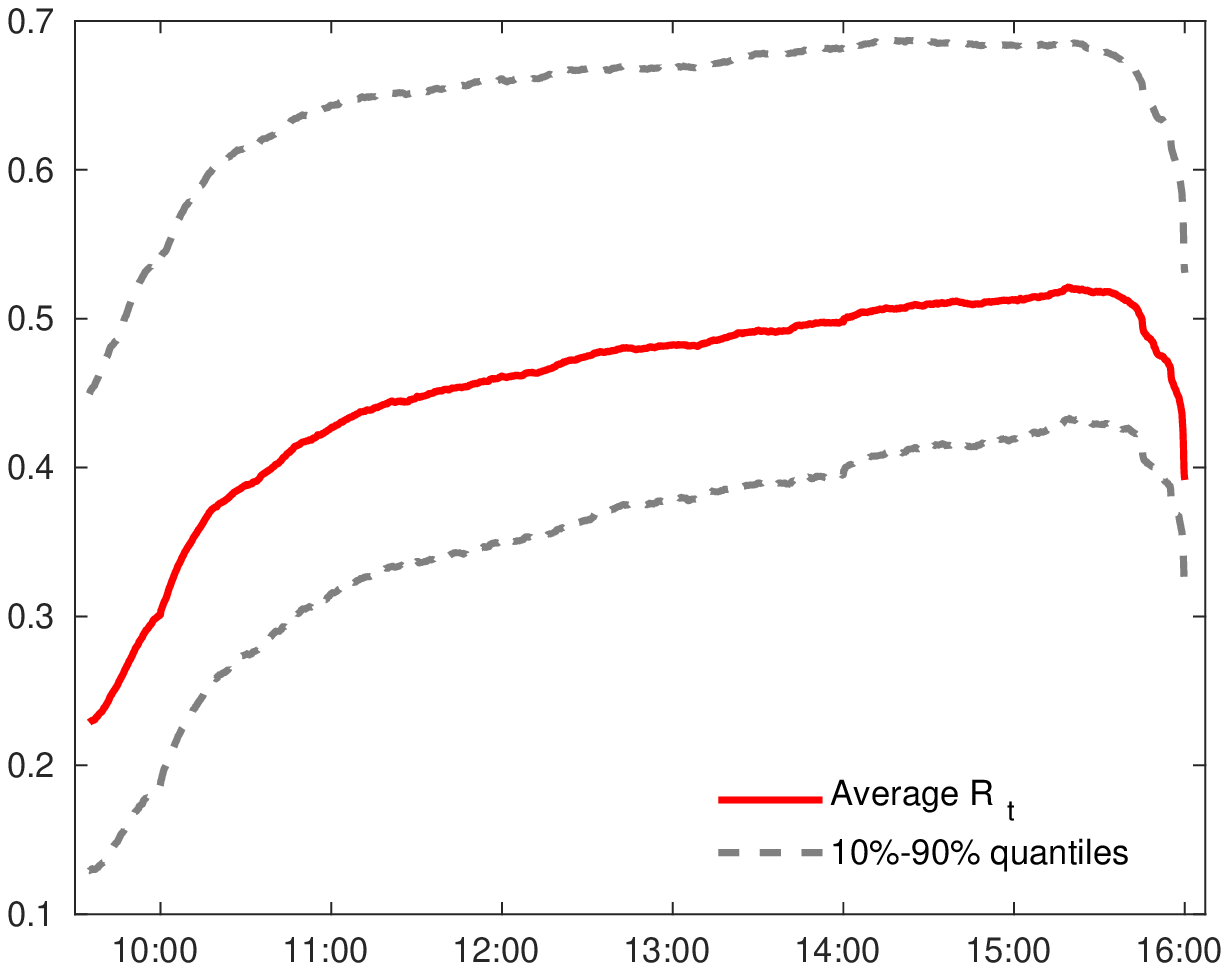} 
  \end{minipage} 
  \caption{Average intraday patterns of the estimated time-varying parameters $D_t$, $H_t^{1/2}$, $\delta_t$ and $R_t$. For each $t=1,\dots,23400$, we report 10\% and 90\% quantiles of the distribution (over assets or couples of assets) of averages of $D_t$, $H_t^{1/2}$, $\delta_t$ and $R_t$ (over days).}
  \label{figTvAsset} 
\end{figure}

We now study in more detail the estimates provided by the local-level model with the aim to extract meaningful information on the behavior of intraday covariances. To gain insights on the level of heterogeneity of correlations, we perform the analysis using the parameterization based on hyperspherical coordinates. We first examine the variation of intraday patterns over time. To this purpose, we take averages of the estimated time-varying parameters over assets or couples of assets. For $t=1,\dots,23400$ and $j=1,\dots,251$, we compute:
\bigskip
\begin{align*}
\tilde{d}_t^j = \frac{1}{n}\sum_{i=1}^n D_t^j(i,i),\quad \tilde{h}_t^j = \frac{1}{n}\sum_{i=1}^n \sqrt{H_t^j(i,i)}\\
 \tilde{\delta}_t = \frac{1}{n}\sum_{i=1}^n \frac{D_t^j(i,i)^2}{H_t^j(i,i)},\quad \tilde{\rho}_t^j = \frac{2}{n(n-1)}\sum_{p>q} R_t^j(p,q)
\end{align*}
where $n=10$. Similarly, to examine the variation of intraday patterns among different assets, we take averages of time-varying parameters over time. For $t=1,\dots,23400$ and $i,p=1,\dots,10$, $q<p$ we compute:
\bigskip
\begin{align*}
\bar{d}_t^i = \frac{1}{N}\sum_{j=1}^N D_t^j(i,i),\quad \bar{h}_t^i = \frac{1}{N}\sum_{j=1}^N \sqrt{H_t^j(i,i)}, \quad\\
 \bar{\delta}_t^i = \frac{1}{N}\sum_{j=1}^N  \frac{D_t^j(i,i)^2}{H_t^j(i,i)},\quad \bar{\rho}_t^{p,q} = \frac{1}{N}\sum_{j=1}^N R_t^j(p,q)
\end{align*}
where $N=251$. The average intraday pattern can be computed by averaging the variables labeled with a ``tilde'' over $j=1,\dots,251$ or, equivalently, by averaging the variables labeled with a bar over $i,p=1,\dots,n$ and $q<p$. For each timestamps $t=1,\dots,23400$, figures \ref{fig:tvTime} and \ref{figTvAsset} show 10\%, 90\% quantiles of variables denoted by ``tilde'' and 10\%, 90\% quantiles of variables denoted by bars, respectively. We also report in each figure the average intraday pattern of time-varying parameters.

The average intraday pattern of volatilities in $D_t$ exhibits the well known U-shape. Volatilities are large at the beginning of the trading day. They gradually decline until the last few minutes, when they steeply increase. The average intraday pattern of microstructure noise volatility has two regimes: a steep decline from 9:30 to 10:00 and a slow decline from 10:00 until the end of the trading day. This dynamic behavior is close to the typical intraday pattern of bid-ask spreads (see e.g. \citealt{McInishWood}), confirming that the local-level is consistently setting apart the efficient log-price process from microstructure effects. The average signal-to-noise ratio exhibits an U-shape as well: it is larger at the beginning of the day ($\delta\sim 10$) and at the end ($\delta\sim 3$), while it is lower in the central part of the day ($\delta\sim 0.5$), suggesting that prices are significantly affected by microstructure effects during a large portion of the trading day.

Correlations exhibit an interesting increasing pattern. At the beginning of the trading day they are low, implying that the dynamics of prices are largely affected by idiosyncratic risk. We then observe a steep increase until 11:00, which is associated with the fast decline of the volatilities. After 11:00, correlations increase, though at a lower rate, until 15:45. At that time, we observe a decline of all the correlations, which corresponds to the increase of volatilities occurring during the last minutes of the trading day. 

\begin{figure}[ht] 
\centering
    \includegraphics[width=0.8\linewidth]{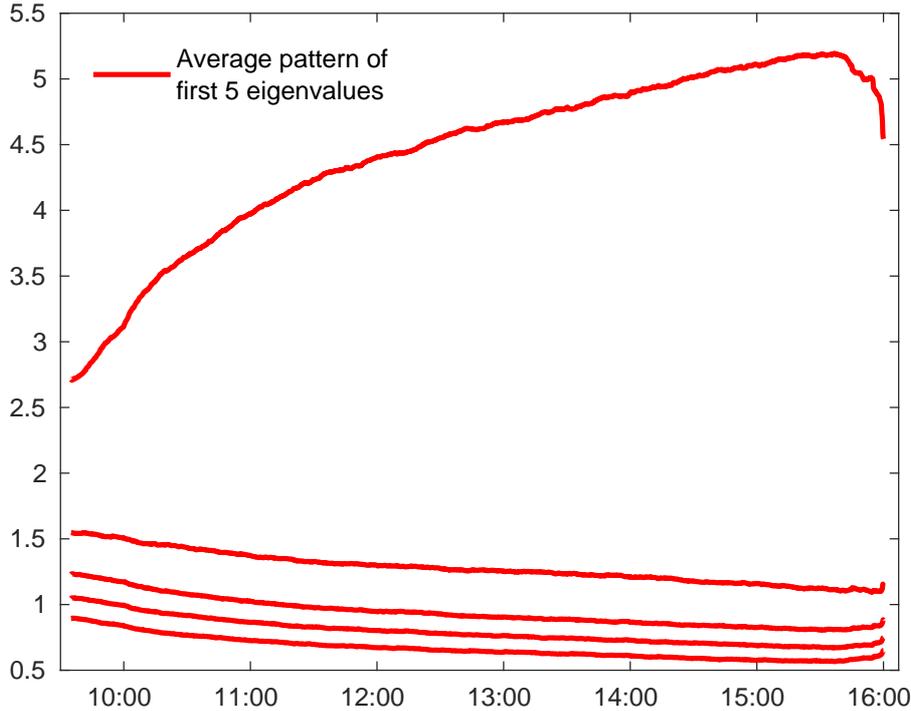} 
    \caption{Average intraday pattern of the first five eigenvalues of the correlation matrix $R_t$.}
  \label{fig:Eigs} 
\end{figure}

Figure \ref{fig:Eigs} shows the average intraday pattern of the first five eigenvalues of the correlation matrix $R_t$. We note that the explanatory power of the first eigenvalue, associated to the market factor, progressively increases during the day. At 15:45, it accounts for $\sim 50\%$ of the total variance. This implies that, while at the beginning of the day asset dynamics are dominated by idiosyncratic risk, a systematic component emerges in the second part of the day. This systematic component is associated to market risk, as all the remaining eigenvalues decrease with time. These results are in agreement with the empirical findings of \cite{AllezBouchaud}, who employed standard sample correlations to study the intraday evolution of dependencies among asset prices. An increase of correlations during the trading day was also found by \cite{BibingerSC} and \cite{KoopmanLitLucasOpschoor}.

\begin{figure}[ht] 
\centering
  \begin{minipage}{0.5\textwidth}
  \centering
    \includegraphics[width=1\linewidth]{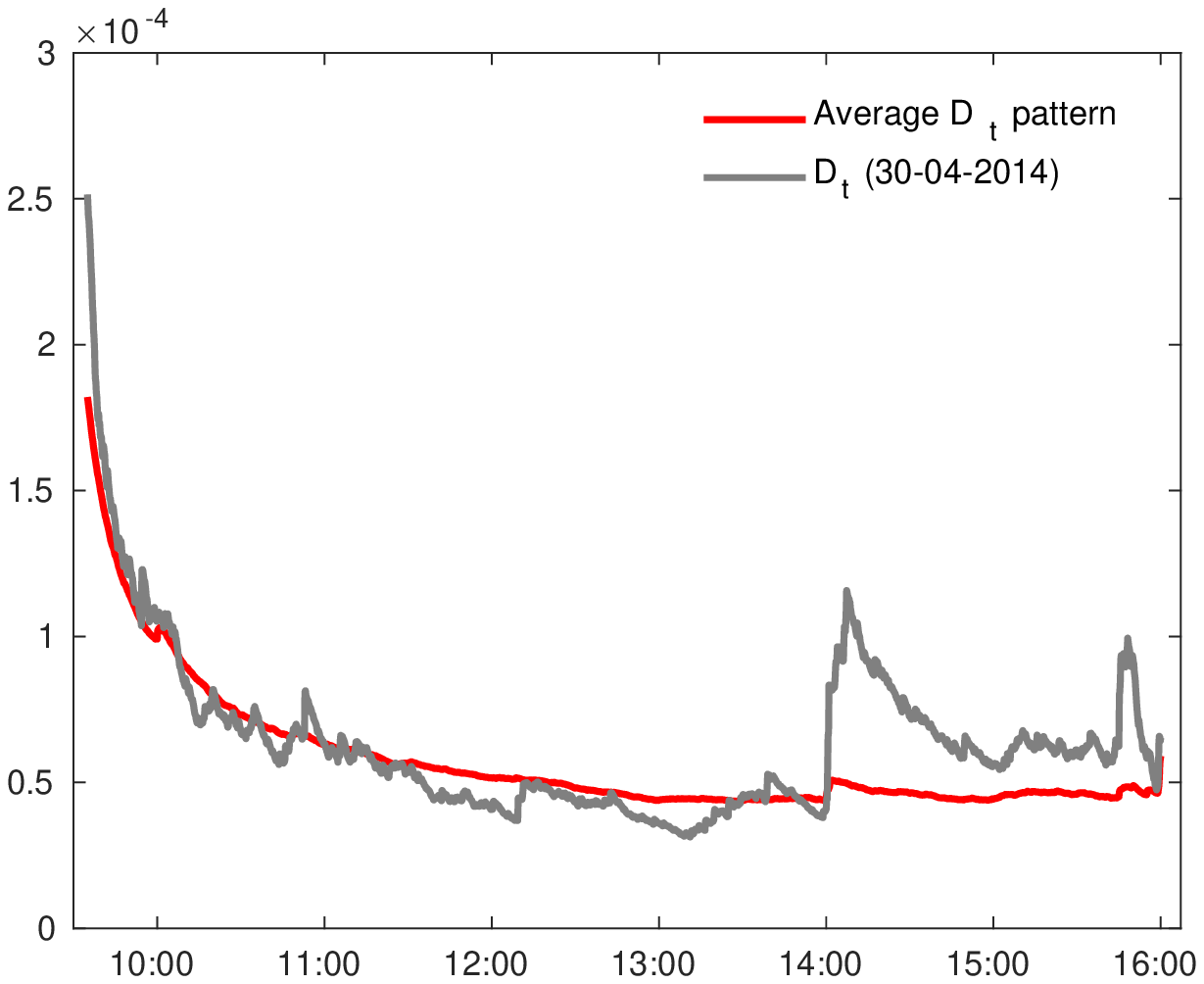} 
  \end{minipage}%
  \begin{minipage}{0.5\textwidth}
  \centering
    \includegraphics[width=1\linewidth]{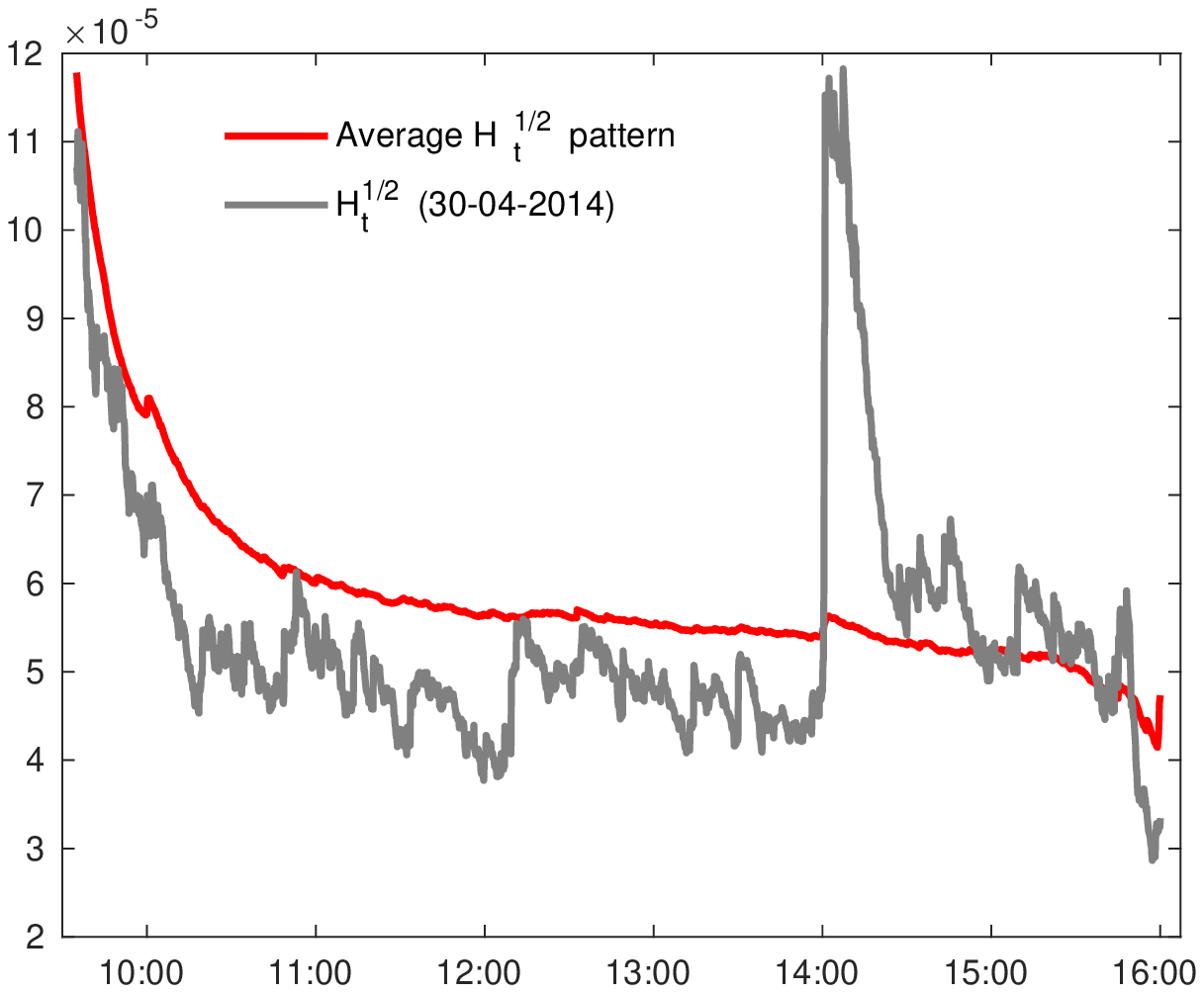} 
  \end{minipage}
  \begin{minipage}{0.5\textwidth}
  \centering
    \includegraphics[width=1\linewidth]{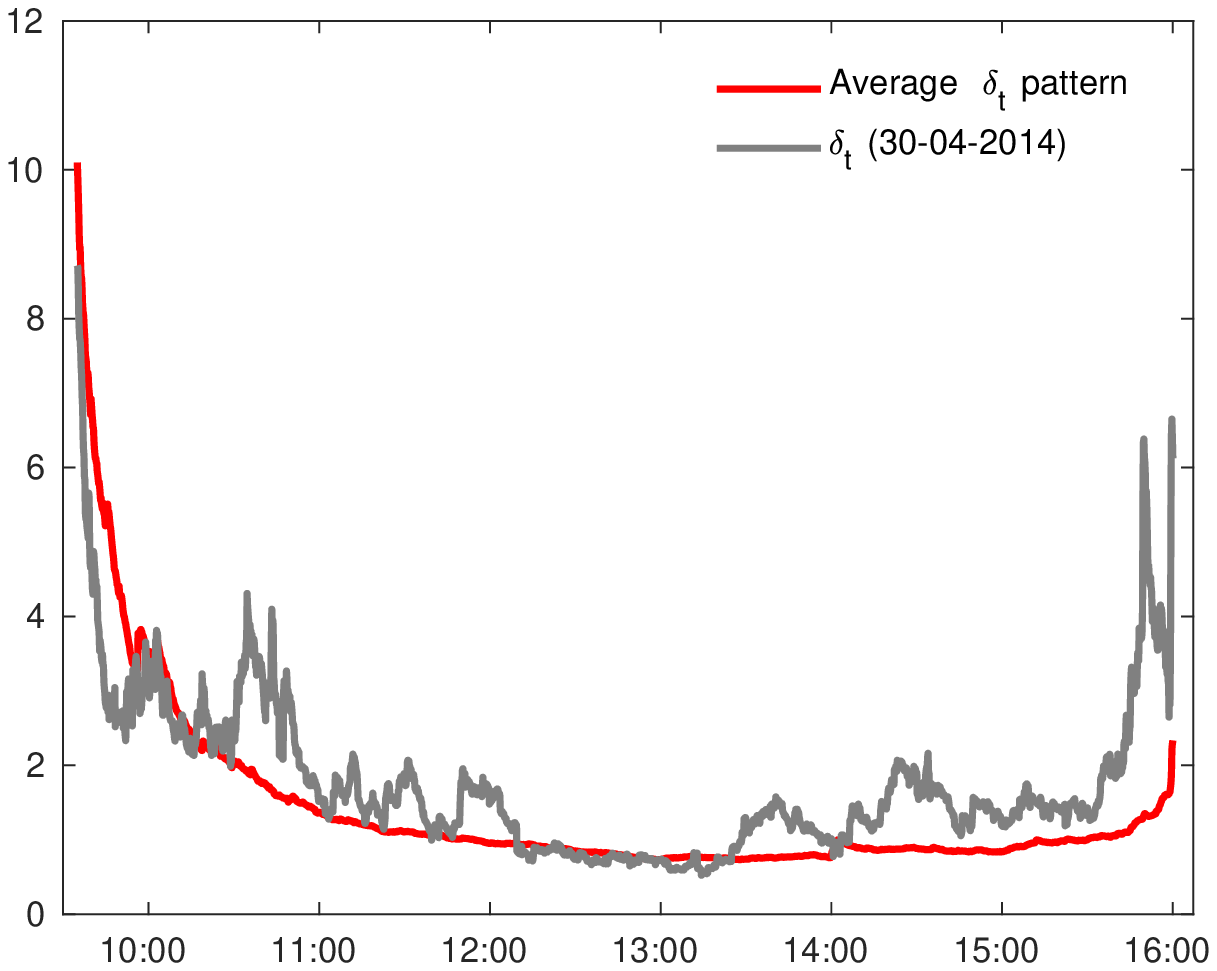} 
  \end{minipage}%
  \begin{minipage}{0.5\textwidth}
  \centering
    \includegraphics[width=1\linewidth]{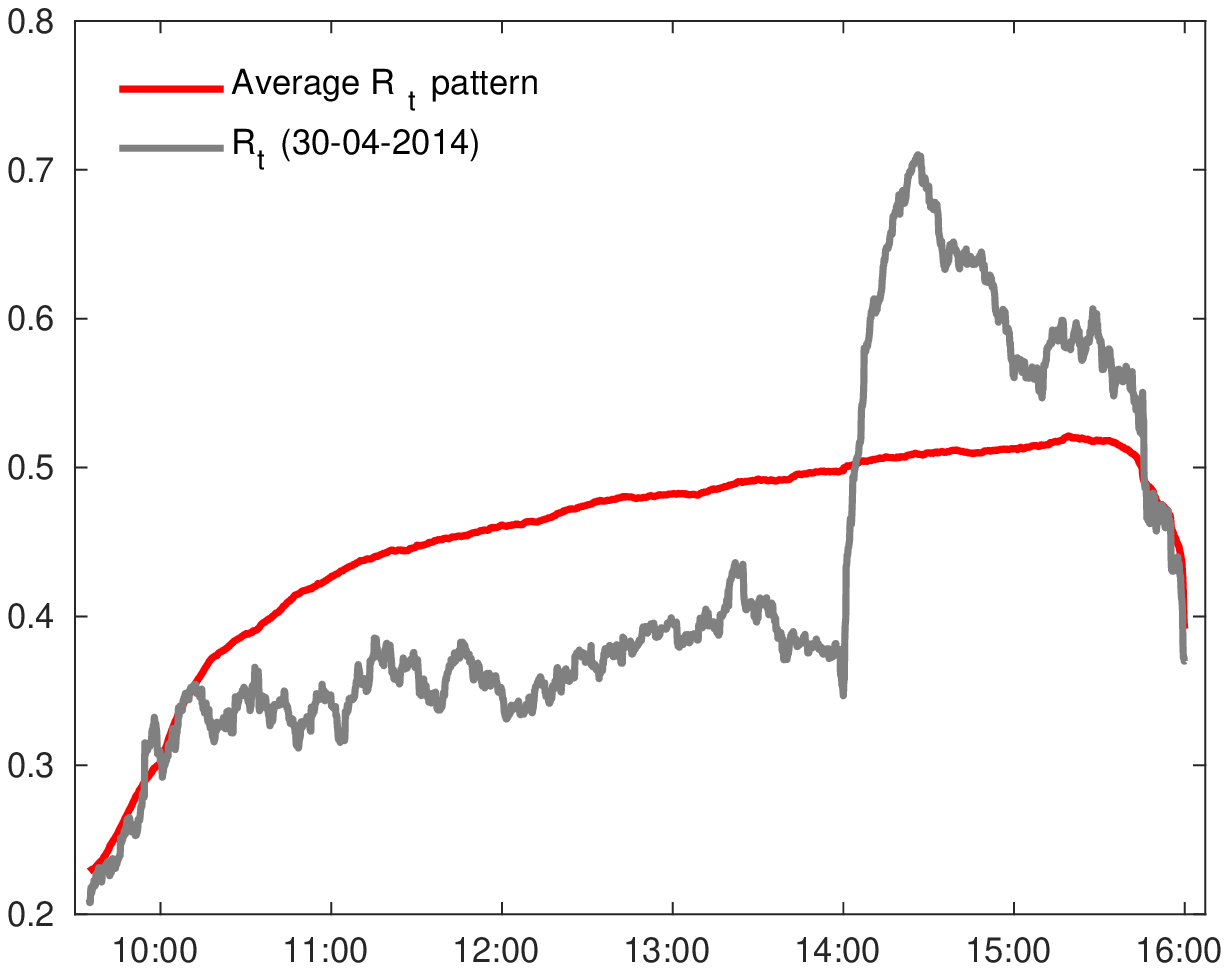} 
  \end{minipage} 
  \caption{We plot $\tilde{d}_t^j$, $\tilde{h}_t^j$, $\tilde{\delta}_t^j$, $\tilde{\rho}_t^j$ computed for $j=82$, corresponding to 30-04-2014}
  \label{figFomc04} 
\end{figure}

\begin{figure}[ht] 
\centering
  \centering
  \begin{minipage}{0.5\textwidth}
  \centering
    \includegraphics[width=1\linewidth]{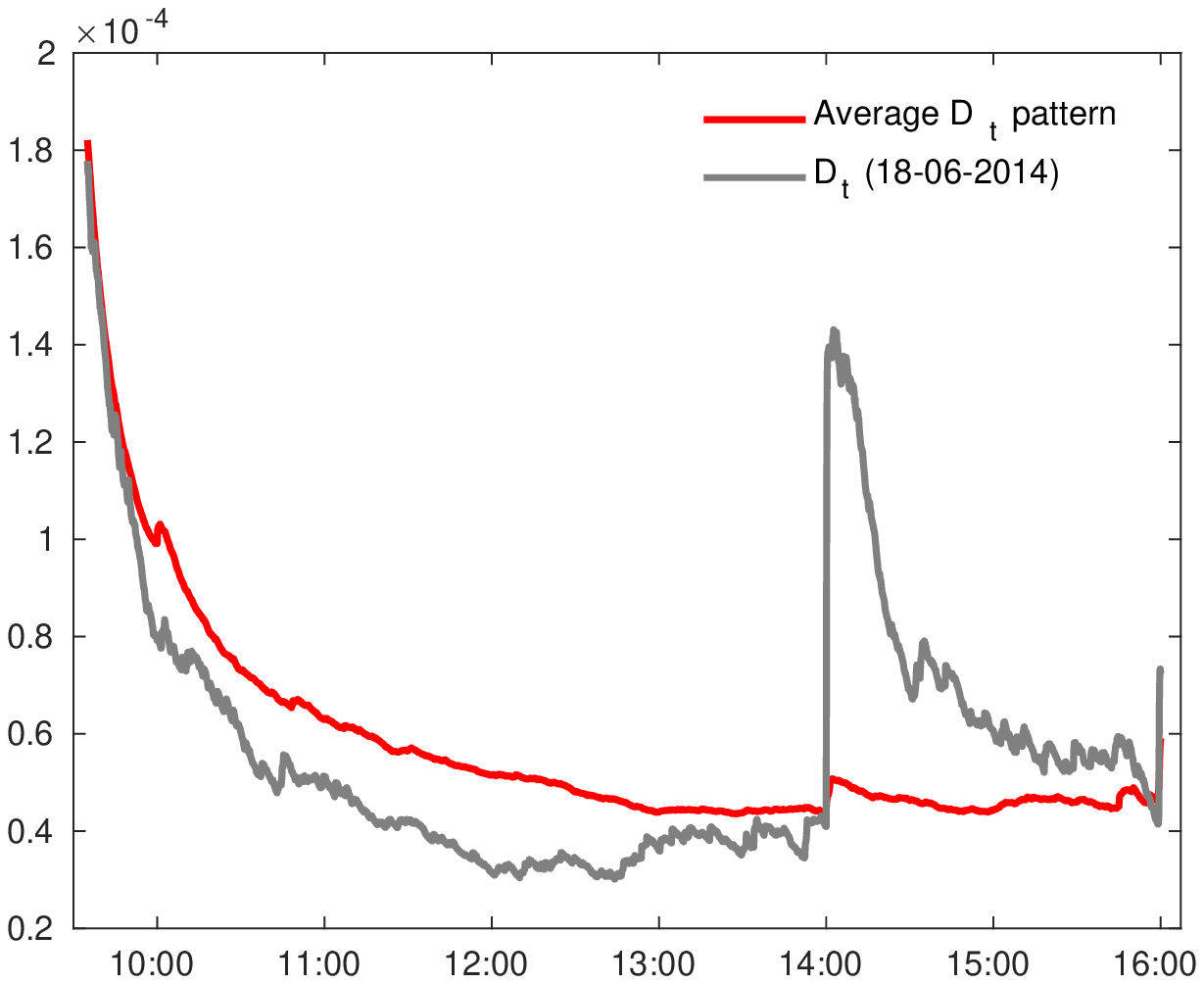} 
  \end{minipage}%
  \begin{minipage}{0.5\textwidth}
  \centering
    \includegraphics[width=1\linewidth]{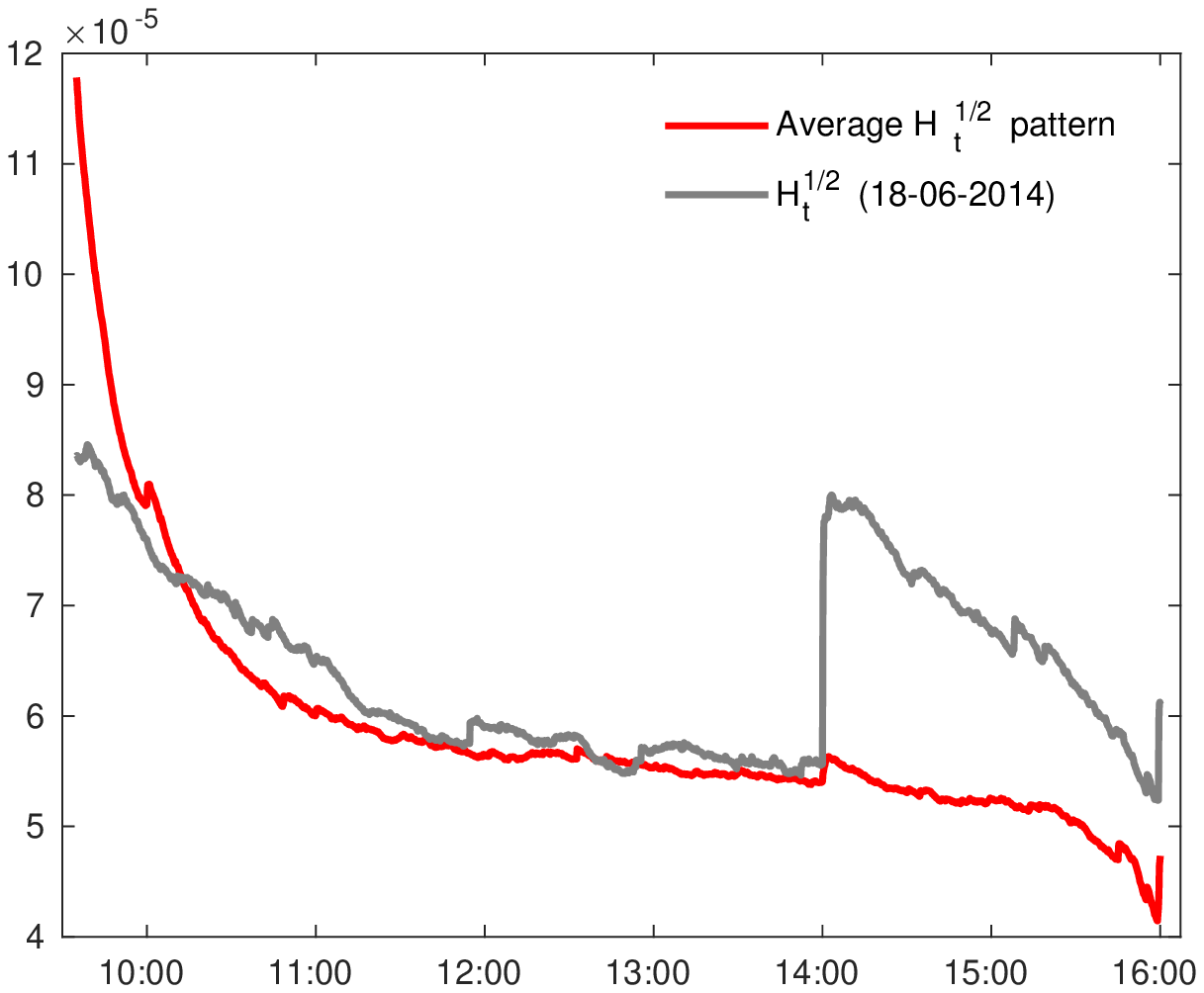} 
  \end{minipage}
  \begin{minipage}{0.5\textwidth}
  \centering
    \includegraphics[width=1\linewidth]{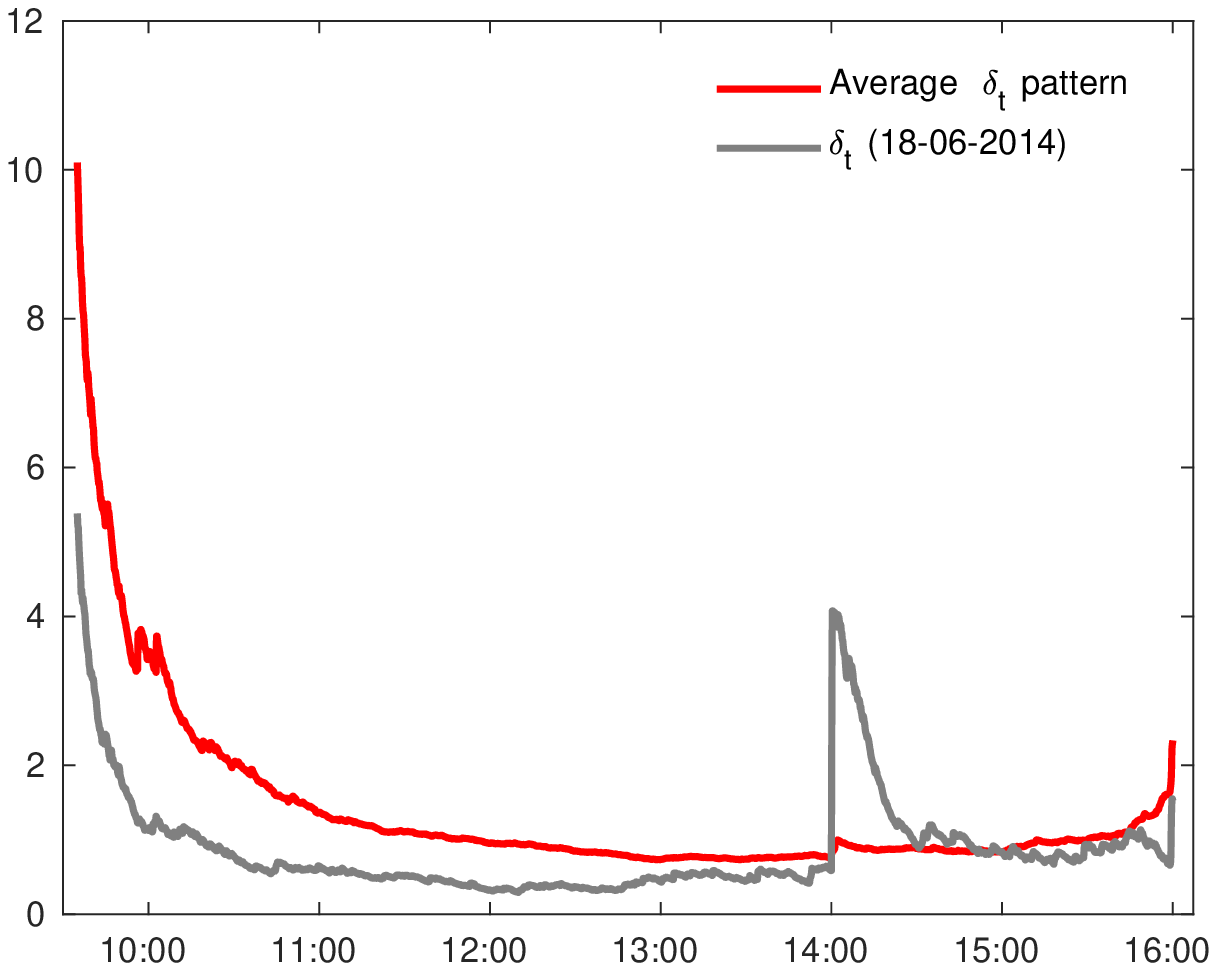} 
  \end{minipage}%
  \centering
  \begin{minipage}{0.5\textwidth}
  \centering
    \includegraphics[width=1\linewidth]{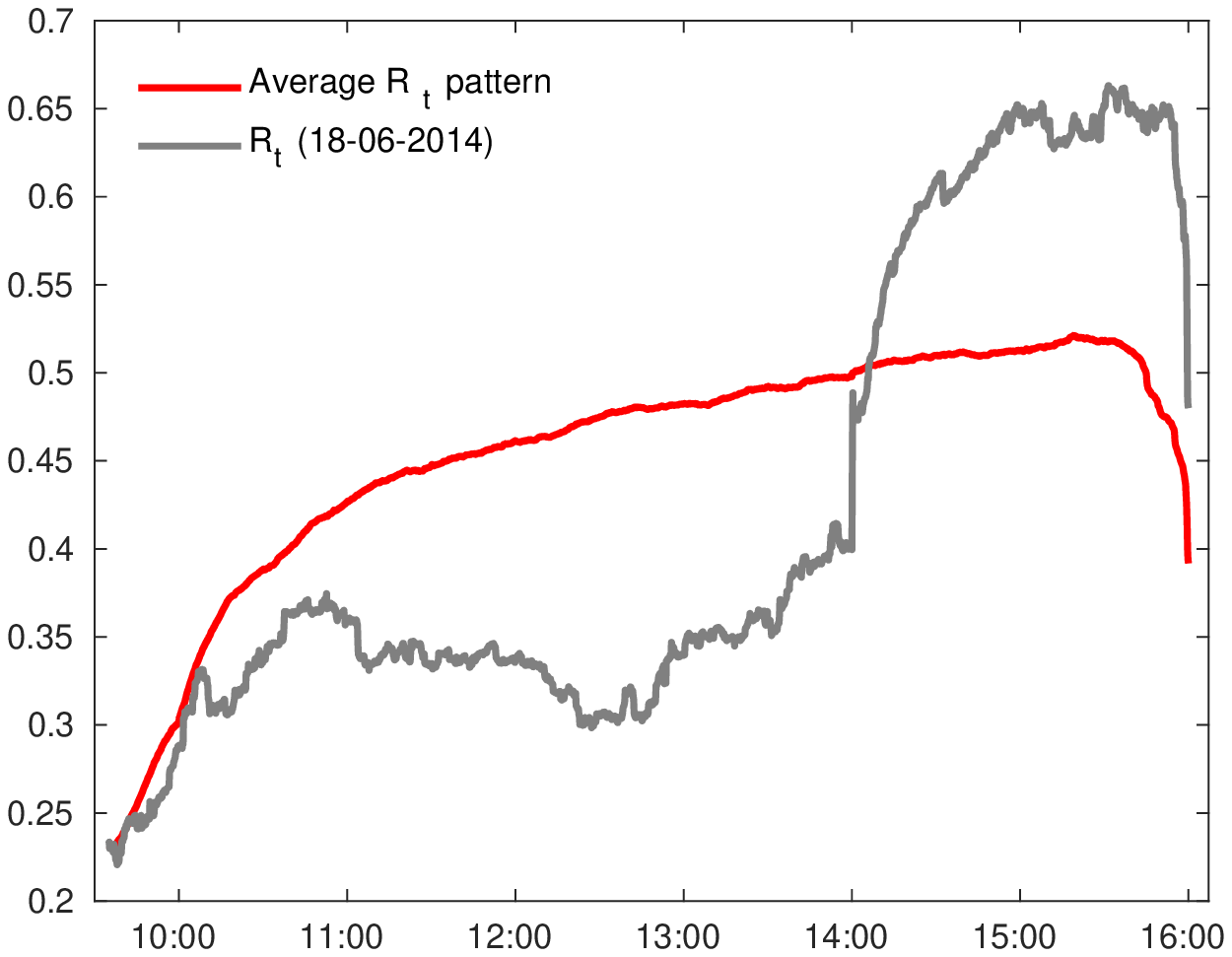} 
  \end{minipage} 
   \caption{We plot $\tilde{d}_t^j$, $\tilde{h}_t^j$, $\tilde{\delta}_t^j$, $\tilde{\rho}_t^j$ computed for $j=116$, corresponding to 18-06-2014}
   \label{figFomc06} 
\end{figure}

Comparing figures \ref{fig:tvTime} and \ref{figTvAsset}, we note that both the U-shape of efficient returns volatility and the decreasing pattern of noise volatility is common to all assets and is stable in time. Similarly, the increase of correlations throughout the day and their sudden drop in the last 15 minutes is common to all the stocks and to all days. There is, however, a remarkable variation on the level of correlations over different days, especially at the end of the trading day. For instance, at 15:30, the 10\% decile of $\tilde{\rho}_t$ in figure \ref{fig:tvTime} is $\sim 0.3$, while the 90\% decile is $\sim 0.7$. From figure \ref{figTvAsset}, we note that even among couples of assets, correlations present a certain degree of heterogeneity. This explains the better in-sample fit found for the hyperspherical coordinates over the equicorrelation parameterization.

While the U-shape and the increasing pattern of correlations are stable in time, there are other patterns which are observed in correspondence of specific events. We plot in figures \ref{figFomc04}, \ref{figFomc06} the quantities $\tilde{d}_t^j$, $\tilde{h}_t^j$, $\tilde{\delta}_t^j$ and $\tilde{\rho}_t^j$ computed in correspondence of two meetings of the Federal Open Market Committee (FOMC) in 2014, the first on 30-04-2014 and the second on 18-06-2014. Compared to the average intraday patterns, we observe significant deviations in the interval between 14:00 and 15:00. This time window coincides with the press conference in which economic news are released by the central bank. The local-level model, exploiting all available 1-second data, instantaneously updates covariances and allows to reconstruct in high resolution the dynamic interaction of covariances with external information. This mechanism shares some similarities with the macroeconomic literature on nowcasting, where dynamic factor models based on Kalman filter with mixed-frequency observations are used to update forecasts of macroeconomic variables (see e.g. \citealt{GIANNONE2008665} and \citealt{DelleMonache2}). Similarly, the local-level model can be employed as a nowcasting tool for high-frequency data. FOMC events are characterized by an increase of volatilities at 14:00, followed by a rapid decline in the subsequent minutes. The increase of microstructure noise volatility at 14:00 is in agreement with the fact that bid-ask spreads are observed to increase during public announcements. After 15:00, volatilities get back to their average pattern. Correlations significantly increase at 14:00 and turn back to their average pattern at the end of the trading day. Note that, due to data reduction, standard return models would not provide such high resolution description of covariance dynamics.

\subsection{Intraday out-of-sample portfolio construction}

In this Section we examine the performance of the model as a nowcasting tool for high-frequency data. To this purpose, we construct intraday global minimum variance portfolios and compare the ex-post realized variance of the portfolio constructed through the local-level model with that of portfolios constructed through alternative methods. Intraday risk and portfolio management has became increasingly popular due to high-frequency trading, which today accounts for about 50\% of total trades in the US equity market (see e.g. \citealt{KARMAKAR2018}).    

\begin{table}[!htbp]
\centering
\small
\setlength{\tabcolsep}{12pt}
 \begin{tabular}{cccccc} 
 \hline
 & Local-level (hyper) & Local-level (equi) & $t$-GAS & DCC & EWMA\\ 
\hline
Avg. portfolio variance & $3.0128$  & $3.3098$ & $3.6926$ & $4.5222$ & $4.6060$\\
N. of days in $\mathcal{M}_{10\%}$ (out of 250)  & $209$ & $173$ & $74$  & $39$ & $35$ \\
Avg. $p$-value  & $0.6802$  &  $0.5552$  &  $0.1512$  & $0.0641$ & $0.04922$ \\
\hline
\end{tabular}
\caption{Average variance ($\times 10^5$) of 1-minute global minimum variance portfolios constructed through out-of-sample covariance forecasts of local-level model (with both hyperspherical coordinates and equicorrelations), $t$-GAS and DCC. We also report the number of times each model is included in the MCS at $10\%$ confidence level and the daily average of $p$-values of MCS tests. }
\label{tab:Ptf}
\end{table}

We adopt the following strategy: on each day $j=2,\dots,251$, we build a sequence of 1-minute portfolios using parameter estimates recovered on day $j-1$. We rebalance the portfolios every minute, ending up with 390 portfolios per day. Following \cite{EngleColacito} and \cite{Patton2009}, we choose as ``best covariance estimator'' the one minimizing the ex-post portfolio variance, computed as:
\begin{equation}
 \sigma_j^2 = \sum_{k=1}^{390}(\hat{w}_{k,j}'r_{k,j})^2
\end{equation}
where $\hat{w}_{k,j}$ denotes the solution of the global minimum variance problem at minute $k$ of day $j$ and $r_{k,j}$ is the vector of 1-minute returns at minute $k$ of day $j$. We employ the same dataset of $n=10$ assets used in the previous analysis. As benchmarks, we consider the $t$-GAS model, the DCC and the EWMA. The latter two are estimated by synchronizing prices at the frequency of one minute, while the $t$-GAS is estimated through the missing value approach. The ex-post realized portfolio variances provided by the different methods are compared through the Model Confidence Set (MCS) of \cite{mcs}. Specifically, we perform a MCS analysis on each day of the dataset.

Table \ref{tab:Ptf} reports the results of the analysis. The first line shows the average of $\sigma_j^2$, over $j=2,\dots,251$. The local-level model based on hyperspherical coordinates provides lower average variance, followed by the local-level based on equicorrelations and the $t$-GAS. The average variance of DCC and EWMA is significantly larger. The second line shows the number of days each model is included in the MCS at 10\% confidence level. The local-level model based on hyperspherical coordinates is included 209 times, while the one based on equicorrelations is included 173 times. This result confirms what we found in the in-sample analysis, namely that hyperspherical coordinates provide lower average losses, but the deterioration due to the equicorrelation assumption is not excessive. As a consequence of large data reduction and microstructure effects, the $t$-GAS, the DCC and the EWMA are only included $74$, $39$ and $35$ times, respectively. Finally, we report on the third line the average $p$-values of MCS tests, which further corroborates the previous findings.

\section{Conclusions}
\label{sec:conclusions}

In this paper we analyzed the problem of intraday covariance modelling with noisy and asynchronous prices and proposed a modelling strategy that allows to handle these two effects. Specifically, we proposed to model intraday data through a local-level model with time-varying covariance matrices. The dynamics of covariances are driven by the score of the conditional density, allowing to estimate the model in closed form. Asynchronous trading is treated as a standard missing value problem in state-space models. 

The main advantage of this approach is that we model the covariances of latent efficient returns rather than those of observed returns. In an high-frequency setting, where observations are asynchronous, models for observed returns are subject to large data reduction if a missing value approach is adopted. On the other hand, if data are synchronized through previous-tick or other interpolation schemes, a large number of ``artificial'' zero returns are introduced, which lead to a downward bias on correlations. Another advantage is that the estimated covariances are robust to microstructure effects, as the model includes an additive measurement error term. 

In our simulation analysis we studied the finite sample properties of the maximum-likelihood estimator and found that it remains unbiased in scenarios characterized by high levels of asynchronicity. We then showed the advantages of the proposed methodology over standard correlation models for observed returns. In particular, in presence of missing values, the use of the methodology is preferable even under a misspecified data generating process with a fat-tail conditional density. The reason lies in the ability of the model to use all available prices in order to reconstruct the efficient price dynamics and the covariances. In contrast, correctly specified return models are subject to large data reduction and the quality of their estimates rapidly deteriorates with increasing asynchronicity. 

The advantages of the methodology are then assessed on empirical data with 1-second transaction data of 10 NYSE assets. The in-sample analyses based on the AIC reveals that benchmark return models are heavily affected by data reduction and that correlations are biased due to market microstructure effects. The analysis of the intraday covariances shows that, while the first trading hours are dominated by idiosyncratic risk, a market risk factor progressively emerges in the second part of the trading day, with the correlations among all couples of assets increasing until the last few minutes of trading. The fact that all available prices are used to build covariances allows to capture instantaneously fast changes on covariance dynamics due to macro-news announcements, such as those released during FOMC announcements.

Finally, we examined whether the above mentioned advantages are able to produce out-of-sample forecast gains. To this purpose, we designed an intraday out-of-sample portfolio test and compared the ex-post realized variance of competing estimators. We found that, in most of the days in our dataset, the variance of portfolios constructed through the local-level is significantly lower than other portfolio variances. Due to the increasing importance of high-frequency trading in financial exchanges, we believe that the approach is of potential usefulness in a wide range of applications.  

\clearpage


\newpage
\bibliography{biblio}
\bibliographystyle{elsarticle-harv}


\newpage
\appendix
\appendixpageoff
\numberwithin{equation}{section}

\huge{\textbf{Appendix}}
\normalsize
\vspace{0.5cm}

\small

We first introduce the notation. The $n\times n$ identity matrix is denoted as $\mathcal{I}_n$. We use $\otimes$ to denote the Kronecker product between two matrices. The operator $\text{vec}[\cdot]$, applied to an $m\times n$ matrix $A$, stacks the columns of $A$ into an $mn\times 1$ vector. The operator $\text{diag}[\cdot]$ applied to an $n\times n$ matrix stacks its diagonal elements into an $n\times 1$ vector. When applied to an $n\times 1$ vector, it gives a diagonal $n\times n$ matrix with the elements of the vector in the main diagonal. We also introduce the commutation matrix $C_{mn}$, i.e. the $mn\times mn$ matrix such  that $C_{mn}\text{vec}A=\text{vec}A'$ for every $m\times n$ matrix $A$. The derivative of an $m\times n$ matrix function $F(X)$ with respect to the $p\times q$ matrix $X$ is defined as in \cite{AbadirMagnus}, i.e. as the $mn\times pq$ matrix computed as $\partial\text{vec}(F(X))/\partial\text{vec}(X)'$.

\section{Computation of \texorpdfstring{$\dot{v}_t$ and $\dot{F}_t$}{Lg}}
\label{app:kf}

\setcounter{MaxMatrixCols}{20}
\newcommand{\myunderbrace}[2]{\settowidth{\bracewidth}{$#1$}#1\hspace*{-1\bracewidth}\smash{\underbrace{\makebox{\phantom{$#1$}}}_{#2}}}

Let us define $a_t = \text{E}_{t-1}[X_t]$ and $P_t = \text{Cov}_{t-1}[X_t]$. Due to asynchronous trading, $Y_{t}$ is a vector with $n_{t}\le n$ components. We define the $n_{t}\times n$ selection matrix $\Gamma_t$ with ones in the columns corresponding to observed prices. The Kalman filter recursions for the local-level model (\ref{eq:ssm:1}), (\ref{eq:ssm:2}) are given by:
\begin{equation}
\begin{aligned}[c]
v_t &= Y_t-\Gamma_t a_t\\
a_{t+1} & = a_t + K_tv_t
\end{aligned}
\qquad
\begin{aligned}[c]
F_t &= \Gamma_t(P_t+H_t)\Gamma_t' \\
P_{t+1} &= P_t(\mathcal{I}_n-K_t\Gamma_t)' + Q_t
\end{aligned}
\label{eq:KalmanHARK}
\end{equation}
where $K_t=P_t\Gamma_t'F_t^{-1}$. If at time $t$ all observations are missing, we set $a_{t+1}=a_{t}$ and $P_{t+1}=P_t+Q$, as discussed in \cite{DurbinKoopman}. It is convenient to introduce the auxiliary vector of time-varying parameters:
\begin{equation}
 \tilde{f}_t = 
 \begin{pmatrix}
  \text{diag}[H_t] \\
  \text{diag}[D_t^2]\\
  \phi_t
 \end{pmatrix}
\end{equation}
The latter is related to $f_t$ by the following link-function:
\begin{equation}
\tilde{f}_t=\mathcal{L}(f_t)=
 \begin{pmatrix}
  \exp{f_t^{(1)}}\\
  \vdots\\
  \exp{f_t^{(2n)}}\\
  f_t^{(2n+1)}\\
  \vdots\\
  f_t^{(k)}
 \end{pmatrix}
\end{equation}
The Jacobian of the transformation is:
\begin{equation}
J_{\mathcal{L}}=\left(\frac{\partial{\tilde{f_t}}}{\partial{f_t}'}\right)=
\begin{pmatrix}
H_{t}  & 0_{n\times n} & 0_{n\times q}\\
0_{n\times n} & D_t^2 & 0_{n\times q}\\
0_{q\times n} & 0_{q\times n} & \mathcal{I}_{q}
\end{pmatrix}
\label{eq:jabobianSHAR}
\end{equation}  
Note that, using the chain rule, $\nabla_t$ and $\mathcal{I}_{t|t-1}$ can be expressed as:
\begin{equation}
\nabla_t = J_{\mathcal{L}}\tilde{\nabla}_t,\quad \mathcal{I}_{t|t-1} = J_{\mathcal{L}}\tilde{\mathcal{I}}_{t|t-1}J_{\mathcal{L}}
\label{eq:chainRule}
\end{equation}
where:
\begin{equation}
\tilde{\nabla}_t=\left[\frac{\partial{\log p(Y_t|\tilde{f}_{t},\mathcal{F}_{t-1},\Theta)}}{\partial{\tilde{f}_t'}}\right]',\quad \tilde{\mathcal{I}}_{t|t-1}=\text{E}[\tilde{\nabla}_t\tilde{\nabla}_t']
\label{eq:app_shar1}
\end{equation}
which can be computed as in (\ref{eq:nabla}), (\ref{eq:fisher}), but deriving with respect to $\tilde{f}_t$ rather than $f_t$. We thus focus on $\dot{v}_t=\partial v_t/\partial \tilde{f}_t'$ and $\dot{F}_t=\partial\text{vec}(F_t)/\partial \tilde{f}_t'$. As a particular case of the general recursions appearing in \cite{DelleMonache2}, we obtain:
\begin{align}
\dot{v}_t &= -\Gamma_t\dot{a}_t\\
\dot{F}_t &= (\Gamma_t\otimes\Gamma_t)(\dot{P}_t+\dot{H}_t)
\end{align}
where:
\begin{align}
\dot{a}_{t+1} &= \dot{a}_t + (v_t'\otimes\mathcal{I}_n)\dot{K}_t + K_t\dot{v}_t\\
\dot{P}_{t+1} &= \dot{P}_t - (K_t\Gamma_t\otimes\mathcal{I}_n)\dot{P}_t-(\mathcal{I}_n\otimes P_t\Gamma_t')C_{nn_t}\dot{K}_t+\dot{Q}_t\\
\dot{K}_t &= (F_t^{-1}\Gamma_t\otimes \mathcal{I}_n)\dot{P}_t-(F_t^{-1}\otimes K_t)\dot{F}_t\\
\dot{Q}_t &= [(D_tR_t\otimes\mathcal{I}_n) + (\mathcal{I}_n\otimes D_tR_t)]\dot{D}_t + (D_t\otimes D_t)\dot{R}_t
\end{align}

Here $\dot{H}_t = \frac{\partial\text{vec}(H_t)}{\partial \tilde{f}_t'}$, $\dot{D}_t = \frac{\partial\text{vec}(D_t)}{\partial \tilde{f}_t'}$ are $n^2\times k$ matrices given by:
\begin{equation}
  \dot{H}_t = 
 \begin{pmatrix}
1\quad 0\quad  \dots\quad 0 & 0\quad  \dots\quad  0 \\
\vdots  & \vdots  \\
0\quad 1\quad \dots\quad 0 & 0\quad \dots \quad0 \\
\vdots  &\vdots  \\
\vdots &  \vdots   \\
\underbrace{0\quad  \dots\quad  0\quad 1}_{n} &  \underbrace{0\quad \dots\quad 0}_{n+q} 
 \end{pmatrix}
\end{equation}

\begin{equation}
  \dot{D}_t =\frac{1}{2} 
 \begin{pmatrix}
0\quad \dots \quad 0 & \frac{1}{D_{t,11}} \quad 0 \quad \dots\quad  0 & 0\quad \dots\quad  0 \\
\vdots  &  \vdots  & \vdots\\
0\quad \dots\quad 0  &  0\quad \frac{1}{D_{t,22}}\quad  \dots\quad  0 & 0\quad \dots\quad 0\\
\vdots  &  \vdots  & \vdots\\
\vdots  &  \vdots  & \vdots\\
\underbrace{0\quad \dots \quad 0}_{n} & \underbrace{0\quad \dots\dots\quad \frac{1}{D_{t,nn}}}_{n} & \underbrace{0\quad  \dots\quad 0}_{q}
 \end{pmatrix}
\end{equation}

The computation of $\dot{R}_t$ depends on the parameterization. We distinguish the case where the hyperspherical coordinates are used and the case where the equicorrelation parameterization is used. In the first case, we have:
\begin{equation}
 \dot{R}_t = [(Z_t'\otimes \mathcal{I}_n)C_{nn} + (\mathcal{I}_n\otimes Z_t')]\dot{Z}_t
\end{equation}
The derivative of the element $Z_{ij,t}$ with respect to the hyperspherical angle $\theta_{lm,t}$ is given by:
\begin{equation}
\frac{\partial Z_{ij}}{\partial\theta_{lm}} =
\begin{cases} 0 & i>j, j\ne m, l\ge m, l>i \\
              - Z_{ij}\tan\theta_{ij} & i<j, l=i \\
              \frac{Z_{ij}}{\tan\theta_{ij}} & i\le j, l<i
\end{cases}
\end{equation}
Note that the time index was suppressed for ease of notation. In the second case we have:
\begin{equation}
 \dot{R}_t = [0_{n^2\times 2n},\dot{\rho}_t\text{vec}(-\mathcal{I}_n+\mathcal{J}_n)]
\end{equation}
where:
\begin{equation}
 \dot{\rho}_t = \frac{1}{2}\left(1+\frac{1}{n-1}\right)\frac{1}{\text{cosh}^2 \theta_t}
\end{equation}

\end{document}